\documentclass{AIMS}
\usepackage{amsmath}
 \usepackage{paralist}
   \usepackage{graphics} 
  \usepackage{epsfig}
  \usepackage{graphicx}  \usepackage{epstopdf}
 \usepackage[colorlinks=true]{hyperref}
\hypersetup{urlcolor=blue, citecolor=red}

\def\currentvolume{}
 \def\currentissue{}
  \def\currentyear{}
   \def\currentmonth{}
    \def\ppages{}
     \def\DOI{}

\usepackage{bm}
\usepackage{subfigure}
\usepackage{graphicx}
\usepackage{notes2bib}
\usepackage{amsmath}
\usepackage{amssymb}
\usepackage{url}
\usepackage{enumerate}
\usepackage{epstopdf}
\usepackage{color}
\usepackage{mathtools}          
\usepackage{setspace}

\usepackage[usenames,dvipsnames,svgnames,table]{xcolor}
\usepackage{color}
\newcommand{\blue}{\textcolor{blue}}

\newcommand{\red}{\textcolor{red}}

\newcommand{\allblack}{\color{black}{}}

\newcommand{\mb}{\mathbf}
\newcommand{\mm}{\mathbb}
\newcommand{\torus}{{\mathbb{T}}}

\usepackage[mathlines,pagewise,displaymath]{lineno}
\usepackage{lineno}
\newcommand*\patchAmsMathEnvironmentForLineno[1]{
  \expandafter\let\csname old#1\expandafter\endcsname\csname #1\endcsname
  \expandafter\let\csname oldend#1\expandafter\endcsname\csname end#1\endcsname
  \renewenvironment{#1}
     {\linenomath\csname old#1\endcsname}
     {\csname oldend#1\endcsname\endlinenomath}}
\newcommand*\patchBothAmsMathEnvironmentsForLineno[1]{
  \patchAmsMathEnvironmentForLineno{#1}
  \patchAmsMathEnvironmentForLineno{#1*}}
\AtBeginDocument{
\patchBothAmsMathEnvironmentsForLineno{equation}
\patchBothAmsMathEnvironmentsForLineno{align}
\patchBothAmsMathEnvironmentsForLineno{flalign}
\patchBothAmsMathEnvironmentsForLineno{alignat}
\patchBothAmsMathEnvironmentsForLineno{gather}
\patchBothAmsMathEnvironmentsForLineno{multline}
}

\title[Machine-learning construction of a model for a fluid variable] 
      {Machine-learning construction of a model for a macroscopic fluid variable using the delay-coordinate of a scalar observable}

\author[K. Nakai and Y. Saiki]{}


 \email{knakai@ms.u-tokyo.ac.jp}
 \email{yoshi.saiki@r.hit-u.ac.jp}

\begin{document}
\maketitle
\centerline{\scshape Kengo Nakai}
\medskip
{\footnotesize
 \centerline{Graduate School of  Mathematical Sciences, The University of Tokyo}
   \centerline{3-8-1 Komaba, Tokyo 153-0041, Japan}
} 

\medskip

\centerline{\scshape Yoshitaka Saiki}
\medskip
{\footnotesize
 \centerline{Graduate School of Business Administration, Hitotsubashi University}
  \centerline{2-1 Naka, Kunitachi, Tokyo 186-8601, Japan}
   \centerline{and}
 \centerline{JST PRESTO, 4-1-8 Honcho, Kawaguchi-shi, Saitama 332-0012, Japan}
  \centerline{and}
 \centerline{Institute for Physical Science and Technology, University of Maryland}
 \centerline{College Park, MD 20742, USA}
 }

\begin{abstract}
We construct a data-driven dynamical system model for a macroscopic variable the Reynolds number of a high-dimensionally chaotic fluid flow by training its scalar time-series data.
We use a machine-learning approach, the reservoir computing for the construction of the model, and do not use 
the knowledge of a physical process of fluid dynamics in its procedure.
It is confirmed that an inferred time-series obtained from the model approximates the actual one 
and that some characteristics of the chaotic invariant set mimic the actual ones.
We investigate the appropriate choice of the delay-coordinate, especially the delay-time and the dimension,   
which enables us to construct a model having a relatively high-dimensional attractor easily. 
\end{abstract}

\section{Introduction}
Reservoir computing is a brain-inspired machine-learning technique that employs a data-driven dynamical system.
The framework was proposed as Echo-State Networks~\cite{Jaeger_2001,Jaeger_2004} and Liquid-State Machines~\cite{Maass_2002}, and it has been found to be effective in the inference of a future such as time-series, frequency spectra 
and the Lyapunov spectra~\cite{Antonik_2018,Ibanez_2018,Inubushi_2017, Zhixin_2017,Pathak_2018, Pathak_2017,Verstraeten_2007}.

A reservoir is a recurrent neural network whose internal parameters are not adjusted to fit the data in the training process. 
Only an output layer is trained.
Therefore, 
the total computational costs are relatively low in comparison 
with many other machine learning techniques having the same dimensional neural networks. 
Many physical phenomena including a fluid flow are deterministic, and thus can be described by a high-dimensional dynamical system, even thought they have a complex behavior.
That is why the reservoir computing with a high-dimensional neural networks can be useful for the construction of a model for such a phenomenon. 

In our previous paper~\cite{nakai_2018}, we infer both microscopic and macroscopic behaviors of a three-dimensional chaotic fluid flow using reservoir computing.
We presented two ways of inference of the complex behavior: the first, called partial inference, requires continued knowledge of partial time-series data during the inference as well as past time-series data, while the second, called full inference, requires only past time-series data as training data. For the first case, we are able to infer long-time motion of microscopic fluid variables. For the second case, we showed that the reservoir dynamics constructed from only past data of energy functions can infer the future behavior of energy functions and reproduced the energy spectrum.

In various experiments and observations of high-dimensional complex phenomena, 
there are usually much smaller number of measurements than the Lyapunov dimensions of the attractor.
Even in such cases, we can efficiently 
construct a dynamical model by generating high-dimensional input data $\mb{u}$ for the reservoir computing by using the delay-coordinate~\cite{nakai_2018,Embedology,Takens}. 

The current paper focuses on the model construction and the full-inference of a macroscopic variable, the Taylor microscale Reynolds number, when the scalar time-series is accessible as measurements. 
We evaluate the model in many ways, and discuss details of the appropriate choice of the delay-coordinate created from the single observable.
This will be useful for readers who wish to construct a reservoir model by themselves.

After reviewing the procedure of the reservoir computing in Sec.~\ref{reservoir} and 
the generation of time-series data of a fluid flow in Sec.~\ref{fluid}, 
we show that the constructed reservoir model recovers various properties of a fluid flow obtained from the Navier-Stokes equation in Sec.~\ref{inference}.
We investigate the effective choice of delay-coordinate in order to construct a model  
in Sec.~\ref{delay}.
We summarize our results in Sec.~\ref{summary}.

\section{Reservoir computing}\label{reservoir}
Reservoir computing is recently used in the inference of complex dynamics~\cite{Ibanez_2018,Lu_2018,Zhixin_2017,Pathak_2018, Pathak_2017}. 
It focuses on the determination of a linear function from the reservoir state vector to variables to be inferred 
(see eq.~(\ref{eq:output2})).
Here we review the outline of the method \cite{Jaeger_2004,Zhixin_2017}. 
In this paper, we construct a model dealing with so called full-inference, in which there is no observable data in the inference phase \cite{nakai_2018}.  \\
\indent We consider a dynamical system 
\begin{equation*}
\frac{d\mb{\phi}}{dt}=\mb{f}(\mb{\phi}),
\end{equation*} 
together with a pair of $\phi$-dependent, vector valued variables 
\begin{equation}
\mb{u}=\mb{h}_1(\mb{\phi})\in \mm{R}^{M} 
~~\text{and}~~
\mb{s}=\mb{h}_2(\mb{\phi})\in \mm{R}^{M}.\label{eq:input}
\end{equation} 
We seek a method for using the 
knowledge of $\mb{u}$ to determine an estimate $\Hat{\mb{s}}$ of $\mb{s}$ as a function of time when direct measurement of $\mb{s}$ 
is not available.
We have a knowledge $\mb{u}$ and  $\mb{s}$ during the {\bf training phase} for $t\le T$, 
$\mb{u}$ and  $\mb{s}$  are unknown during the {\bf inference phase} for $t>T$.
Therefore, $\mb{u}$ during the inference phase is replaced by $\Hat{\mb{s}}$ in the previous step. See eq. \eqref{eq:full-reservoir} for the detail. 
\allblack\\
%
\indent The dynamics of the {\bf reservoir state vector} 
\begin{equation*}
\mb{r}\in \mm{R}^{N}~(N \gg M),
\end{equation*}
 is defined by the neural network 
\begin{equation}
	\mb{r}(t+\Delta t)=(1-\alpha)\mb{r}(t)+\alpha \tanh(\mb{A}\mb{r}(t)+\mb{W}_{\text{in}}\mb{u}(t)
	),\label{eq:reservoir}
\end{equation}
where $\Delta t$ is a relatively short time step, 
and 
\begin{equation*}
\tanh(\mb{q})=(\tanh(q_1), \tanh(q_2),\cdots,\tanh(q_N))^{\text{T}},
\end{equation*}
for a vector $\mb{q} = (q_1,q_2,\cdots,q_N)^{\text{T}}$. 
Here,  $\text{T}$ represents  the  transpose of a matrix. 
The matrix $\mb{A}$ is a weighted adjacency matrix, and the $M$-dimensional 
input $\mb{u}$ is fed in to the $N$ reservoir nodes via a linear input weight matrix denoted by $\mb{W}_{\text{in}}$.
The parameter $\alpha$ ($0<\alpha\le 1$) 
adjusts the nonlinearity of the dynamics of $\mb{r}$, 
and is chosen depending upon the complexity of the dynamics of measurements and the time step $\Delta t$. \\
\indent Each row of $\mb{W}_{\text{in}}$ has one nonzero element, chosen from a uniform distribution on $[-\sigma,\sigma]$.
The matrix $\mb{A}$ is chosen from a sparse random 
matrix in which the fraction of nonzero matrix elements is $D/N$, 
so that the average degree of a reservoir node is $D$. 
The $D$ non-zero components are chosen from a uniform distribution on $[-1, 1]$. 
Then we uniformly rescale all the elements of $\mb{A}$ so that the largest value of the magnitudes of its eigenvalues becomes $\rho$. \\
\indent The output, which is a $M$-dimensional vector, is taken to be a linear function of the reservoir state vector $\mb{r}$:
\begin{equation}
	\Hat{\mb{s}}(t)=\mb{W}_{\text{out}}\mb{r}(t)+\mb{c}.\label{eq:output}
\end{equation}
The reservoir state vector $\mb{r}$ evolves following eq.~(\ref{eq:reservoir}) with input 
$\mb{u}(t)$, 
starting from random initial state $\mb{r}(-T_0)$ whose elements are chosen from $(0, 1]$ in order not to diverge, 
where $T_0=L_0\Delta t~(\gg 1)$ is the transient time for $\mb{r}(t)$ ($t>0$) to be on the attractor.
We obtain $L=T/\Delta t$ steps of reservoir state vectors $\{\mb{r}(l\Delta t)\}_{l=1}^{L}$ by iterating eq.~(\ref{eq:reservoir}), 
 while we record the variables $\{\mb{s}(l\Delta t)\}_{l=1}^{L}$ by using  the actual measurements from eq. (\ref{eq:input}) for the training phase. 
\allblack
\\
\indent{\bf Determination of $\mb{W}_\text{out}$ and $\mb{c}$.}
\indent 
We determine $\mb{W}_\text{out}$ and $\mb{c}$ 
so that the reservoir output  $\Hat{\mb{s}}$ (eq. \eqref{eq:output}) approximates the measurement $\mb{s}$ for $0< t \le T$~(training phase), which is a training process in the reservoir computing.
We determine them  by minimizing the following quadratic form with respect to $\mb{W}_\text{out}$ and $\mb{c}$:
\begin{equation}
\displaystyle\sum^{L}_{l=1} \|(\mb{W}_\text{out}\mb{r}(l\Delta t)+\mb{c})-\mb{s}(l\Delta t)\|^2
+\beta[Tr(\mb{W}_\text{out}\mb{W}^{\text{T}}_\text{out})],\label{eq:minimize}
\end{equation}
where $\|\mb{q}\|^2=\mb{q}^{\text{T}} \mb{q}$ for a vector $\mb{q}$,
and the second term is a regularization term introduced to avoid overfitting 
$\mb{W}_\text{out}$ for $\beta \ge 0$.
When the training is successful, $\Hat{\mb{s}}(t)$ should approximate the desired unmeasured quantity $\mb{s}(t)$ for $t>T$~(inference phase). 
Following eq.~(\ref{eq:output}), we obtain 
\begin{equation}
	\Hat{\mb{s}}(t)=\mb{W}^*_\text{out}\mb{r}(t)+\mb{c}^*,  \label{eq:output2}
\end{equation}
where $\mb{W}^*_\text{out}$ and $\mb{c}^*$ denote the solution for the minimizers 
of the quadratic form~(\ref{eq:minimize}):
\begin{align}
	\mb{W}^*_\text{out}&=\delta\mb{S}\delta\mb{R}^{T}(\delta\mb{R}\delta\mb{R}^{T}+\beta\mb{I})^{-1}, \label{eq:wout-c1}\\
	\mb{c}^*&=-[\mb{W}^*_\text{out}\overline{\mb{r}}-\overline{\mb{s}}], \label{eq:wout-c2}
\end{align}
where $\overline{\mb{r}}=\sum^{L}_{l=1} \mb{r}(l \Delta t)/L$, 
$\overline{\mb{s}}=\sum^{L}_{l=1} \mb{s}(l \Delta t)/L$,
and $\mb{I}$ is the $N \times N$ identity matrix, $\delta\mb{R}$ (respectively, $\delta\mb{S}$) is the matrix 
whose $l$-th column is $\mb{r}(l\Delta t)-\overline{\mb{r}}$ (respectively, $\mb{s}(l\Delta t)-\overline{\mb{s}}$)
(see~\cite{Lukosevicius_2009}~P.140 and \cite{Tikhonov_1977} Chapter 1  for details).
\allblack

In the inference phase for $t>T$, eq.(\ref{eq:reservoir}) is written as 
\begin{equation}
\mb{r}(t+\Delta t)=(1-\alpha)\mb{r}(t)+\alpha \tanh(\mb{A}\mb{r}(t)+\mb{W}_{\text{in}}\Hat{\mb{s}}(t)),\label{eq:full-reservoir}
\end{equation}
by setting $\mb{u}(t)$ as $\Hat{\mb{s}}(t)$ obtained from eq.~(\ref{eq:output2}). \\
\indent We define a reservoir model by eqs. (\ref{eq:output2}) and (\ref{eq:full-reservoir}) under the values determined by eqs. (\ref{eq:wout-c1}) and (\ref{eq:wout-c2}) through the training data in a time-interval $[0,T]$.  
\allblack
The main variables and matrices in the reservoir computing are summarized in Table~\ref{tab:variable}.

\indent {\bf Normalization of a variable.}
In order to consider the effect of all the variables equally, we take the normalized value $\tilde{x}(t)$ for each variable $x(t)$, which will be used in the procedure of our reservoir computing:
\begin{equation*}
\tilde{x}(t)=[x(t)-X_1]/X_2,
\end{equation*}
where $X_1$ is the mean value and $X_2$ is the variance. 
When  we reconstruct $x(t)$ in the inference phase from $\tilde{x}(t)$, 
we employ $X_1$ and $X_2$ obtained in the training phase.
 Due to the normalization we can avoid adjustments of $\sigma$. 
 
{\bf Parameter choice.} We apply a method of reservoir computing described above in order to construct a model.
The sets of parameter values used are shown in Table~\ref{tab:parameter}.

\begin{table}[htb]
		\begin{tabular}{|l|l|} 
			\hline  	
           		  \multicolumn{2}{|c|}{variable}\\ \hline
  			~$\mb{u}~(\in \mathbf{R}^M)$&input variable\\ \hline
 			~$\mb{r}~(\in \mathbf{R}^N)$&reservoir state vector\\ \hline
 			~$\mb{s}~(\in \mathbf{R}^M)$&actual output variable obtained from Navier-Stokes equation\\ \hline
 			~$\Hat{\mb{s}}~(\in \mathbf{R}^M)$& inferred output variable obtained from  reservoir computing\\ \hline
 			~$\mb{A}~(\in \mathbf{R}^{N \times N})$&weighted adjacency matrix \\ \hline
 			~$\mb{W}_{\text{in}}~(\in \mathbf{R}^{M \times N})$& linear input weight \\ \hline
 			~$\mb{W}_{\text{out}}~(\in \mathbf{R}^{N \times M})$& matrix used for translation from $\mb{r}$ to output variable  $\Hat{\mb{s}}$ \\ \hline
 			~$\mb{c}~(\in \mathbf{R}^{M})$& vector  used for translation from $\mb{r}$ to output variable  $\Hat{\mb{s}}$ \\ \hline
  			~$\tilde{x}$&normalized variable of ${x}$\\ \hline
 \end{tabular}
 		\caption{{\bf The list of variables and matrices in the reservoir computing.} 
 		}
 		\label{tab:variable}
\end{table}

\begin{table}[htb]
		\begin{tabular}{|l|l|r|r|r|r|} 
			\hline  	
           		  \multicolumn{2}{|c|}{parameter}&Sec.~\ref{inference}&Sec.~\ref{delay}\\ \hline
           		    			~$M$&dimension of input and output variables&14&Table.~\ref{tab:parameter-delay}\\ \hline
	  		~$N$&dimension of reservoir state vector &3000 &2000\\ \hline
 	  		~$D$&parameter of determining $\mb{A}$&120 &80\\ \hline
           		 ~$\Delta t$&time step for reservoir dynamics & \multicolumn{2}{c|}{0.5} \\  \hline
  			~$T_0$&transient time for $\mb{r}$ to be converged&\multicolumn{2}{c|}{3750} \\ \hline
 			~$T$&training time&\multicolumn{2}{c|}{40000} \\ \hline
 			~$L_0${$~(=T_0/\Delta t)$}&number of iterations for the transient&\multicolumn{2}{c|}{7500} \\ \hline
 			~$L$~{$~(=T/\Delta t)$}&number of iterations for the training&\multicolumn{2}{c|}{80000} \\ \hline
	 	 	~$\rho$&maximal eigenvalue of $\mb{A}$ &\multicolumn{2}{c|}{0.7} \\  \hline
   			~$\sigma$&scale of input weights in $\mb{W}_{\text{in}}$&\multicolumn{2}{c|}{0.5} \\ \hline
	   		~$\alpha$&nonlinearity degree of reservoir dynamics&\multicolumn{2}{c|}{0.6}  \\ \hline
 			~$\beta$&regularization parameter &\multicolumn{2}{c|}{0.1} \\ \hline
 \end{tabular}
 		\caption{{\bf The list of parameters and their values used in the reservoir computing in each section.} 
 		}
 		\label{tab:parameter}
\end{table}

 
\section{Generation of a fluid flow data}\label{fluid}
Modelling and inference of a fluid flow are important problems in many areas~\cite{Clark_2018,nakai_2018}.
In this paper, we construct a model for a macroscopic variable of a fluid flow, 
especially the time-dependent ``Taylor microscale Reynolds number'' which reflects the degree of complexity in the fluid flow.
We generate training data by the direct numerical simulation of the Navier-Stokes equation, 
which is also used for the reference data in the inference phase in order to evaluate the constructed reservoir model. 
It should be remarked that the Navier-Stokes equation and its physical property are not considered at all when constructing a reservoir model.

{\bf Generation of training data.}
In order to generate measurements of the reservoir computing, 
we employ the direct numerical simulation of the incompressible three-dimensional Navier-Stokes equation 
under periodic boundary conditions:
	\begin{align*}
  \begin{cases}
		\partial_t v -\nu \Delta v
			+(v \cdot \nabla) v+\nabla \pi=f,~\nabla \cdot v=0, ~\mathbb{T}^3\times(0,\infty),\\
v\big| _{t=0}=v_0\quad \text{with $\nabla \cdot v_0=0$}, ~~~~~~~~~~~~~~~~~~~~\mathbb{T}^3,
  \end{cases}
	\end{align*}
where $\torus=[0,1)$, $\nu>0$ is a viscosity parameter, $\pi (x,t)$ is pressure, and $v(x,t)= (v_1(x,t),v_2(x,t),v_3(x,t))$ is velocity.
Throughout this paper, we set $\nu = 0.058$, under which the fluid flow shows an intermittent behavior between laminar and bursting states. See such a behavior in the bottom panel of Fig.~\ref{fig:full-macro}.  
We use the Fourier spectral method~\cite{ishioka_1999} with $N_0(=9)$ modes in each of three directions, meaning that the system is approximated by 
$2(2 N_0+1)^3~(=13718)$-dimensional ordinary differential equations (ODEs).
The ODEs are integrated by the 4th-order Runge--Kutta scheme, and the forcing is input into the low-frequency variables at each time step so as to preserve the energy of the low-frequency part.
See~\cite{ishioka_1999,nakai_2018} for the details.  

{\bf Reynolds number $R_{\lambda}$.}
We focus on the time-series of the Taylor microscale Reynolds number,
a macroscopic variable representing the degree of complexity of a fluid flow. 
The total energy $E(t)$ is defined by 
\begin{equation*}
E(t)=\sum_{\kappa \in D} \sum_{\zeta=1}^{3} \left(
\mathcal{F}_{[v_{\zeta}]}(\kappa,t)
\right)^2, 
\end{equation*}
where 
\begin{equation*}
	\mathcal{F}_{[v_{\zeta}]}(\kappa,t):= \dfrac{1}{(2\pi)^3} \displaystyle\int_{\mathbb{T}^3}
			v_{\zeta}(x, t)
		e^{-i(\kappa\cdot x)}dx \quad(\zeta = 1,2,3), 
\end{equation*}
and $D=\{(\kappa_1,\kappa_2,\kappa_3) \in  \mathbb{Z}^3 \mid \kappa_1,\kappa_2,\kappa_3 \in [-9,9] \}$.
The Taylor microscale Reynolds number $\check{R}_{\lambda}(t)~$\cite{Ishihara_2003} is defined 
as follows:
	\begin{align*}
		\check{R}_{\lambda}(t) := \dfrac{\sqrt{(2/3)E(t)}\lambda}{\nu}
		 = \sqrt{\dfrac{20 E(t)^2}{3\nu \epsilon(t)}}, 
	\end{align*}
where 
\begin{equation*}
\epsilon(t)=2\nu \sum_{\kappa \in D}  \sum_{\zeta=1}^{3} \left|\kappa\right|^2 \left(
\mathcal{F}_{[v_{\zeta}]}(\kappa,t)
\right)^2,
\end{equation*}
 is the average rate of energy dissipation per unit mass 
and 
\begin{equation*}
\lambda=\left(\frac{15 \nu (2/3)E(t)}{\epsilon(t)}\right)^{1/2},
\end{equation*}
is the characteristic length of a turbulent fluid flow.
The length roughly corresponds to that of an energy input in this study. 

In order to get rid of the high-frequency fluctuation, we take the
short-time average
\begin{equation*}
{R_{\lambda}}(t)=\sum_{l=99}^{0}\check{R}_{\lambda}(t-l \Delta t^*)/100,
\end{equation*}
where $\Delta t^*=0.05$ is the time step of the integration of the Navier-Stokes equation.  
This helps us to obtain essential low-frequency dynamics of a Reynolds number and construct a model with less computational costs with lower dimension $N$ of the reservoir state vectors. 
The averaged Reynolds number ${R_{\lambda}}$ will be called the Reynolds number, 
and the time-series generated by the direct numerical simulation in the inference phase 
will be called the ``actual data''. 
\\


\section{Construction of a model for a macroscopic variable: Reynolds Number}\label{inference}
Using the reservoir computing discussed in Sec.~\ref{reservoir}, we construct a model by training a time-series data 
of the Reynolds number ${R_{\lambda}}$ (see Sec.~\ref{fluid}) that shows an intermittent behavior between 
laminar and bursting states. 
For its purpose a delay-coordinate vector created from a scalar observable is introduced to the input and output variables.


\subsection{Construction}
{\bf Delay-coordinate.}
The choice of variables for the reservoir model is significant.
Here, we introduce an $M$-dimensional delay-coordinate vector of the Reynolds number with a delay-time $\Delta \tau$ 
as input and output variables $\mb{u}(t)=(u_1(t),u_2(t),\cdots,u_M(t))^{\text{T}}$ and $\mb{s}(t)=(s_1(t),s_2(t),\cdots,s_M(t))^{\text{T}}$ in eq.~\eqref{eq:input}, that is, 
\begin{align}
\mb{u}(t)
&=(\tilde{R}_{\lambda}(t), \tilde{R}_{\lambda}(t-\Delta \tau),\cdots,\tilde{R}_{\lambda}(t-(M-1)\Delta \tau))^{\text{T}}, \label{eq:delay}\\
\mb{s}(t)
&=(\tilde{R}_{\lambda}(t), \tilde{R}_{\lambda}(t-\Delta \tau),\cdots,\tilde{R}_{\lambda}(t-(M-1)\Delta \tau))^{\text{T}}. \label{eq:delay2}
\end{align}
The appropriate choice of the {\bf dimension} $M$ and the {\bf delay-time} $\Delta\tau$ of the delay-coordinate will be discussed in Sec.~\ref{delay}.
\\
{\bf Determination of a model.}
Under the parameters listed in 
Table~\ref{tab:parameter} and randomly chosen matrices $\mb{A}$ and $\mb{W}_\text{in}$, 
 we find a candidate of a reservoir model by fixing $\mb{W}^*_\text{out}$ and $\mb{c}^*$  following the procedure 
 explained in Sec~\ref{reservoir}.
 If the candidate passes a certain criteria concerning the short time  inference, the candidate is considered as a model.
 See Sec.~\ref{delay} for the details of the criteria.
%
Remark that although we can use a training data as some delay components of input data when $t-(M-1)\Delta \tau<T$, 
 we do not use any training data in the inference phase. 
Hereafter throughout this section, we choose one of the  models, and fix the corresponding  set of values $\mb{A}$, $\mb{W}_\text{in}$, $\mb{W}^*_\text{out}$ and $\mb{c}^*$. 
\begin{figure}[h]
\includegraphics[width=0.48\columnwidth,height=0.240\columnwidth]{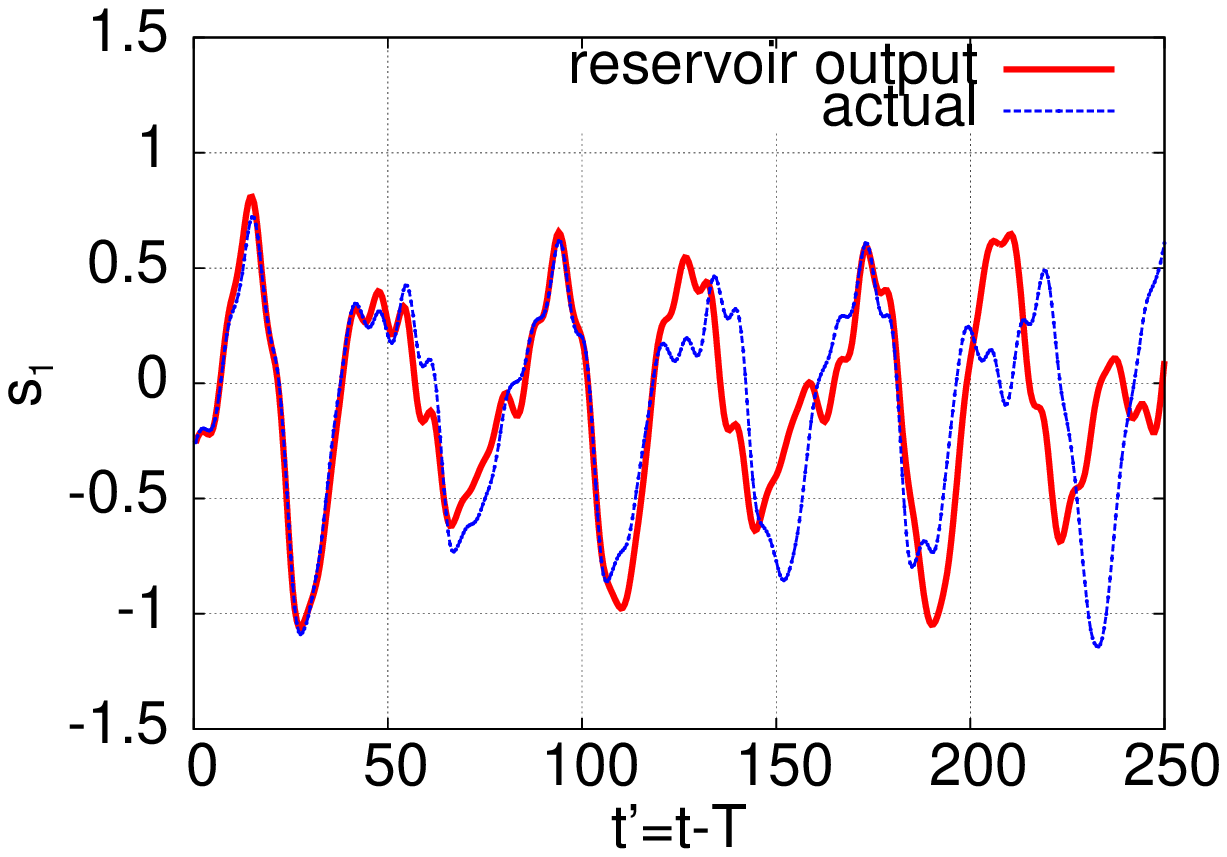}
\includegraphics[width=0.48\columnwidth,height=0.240\columnwidth]{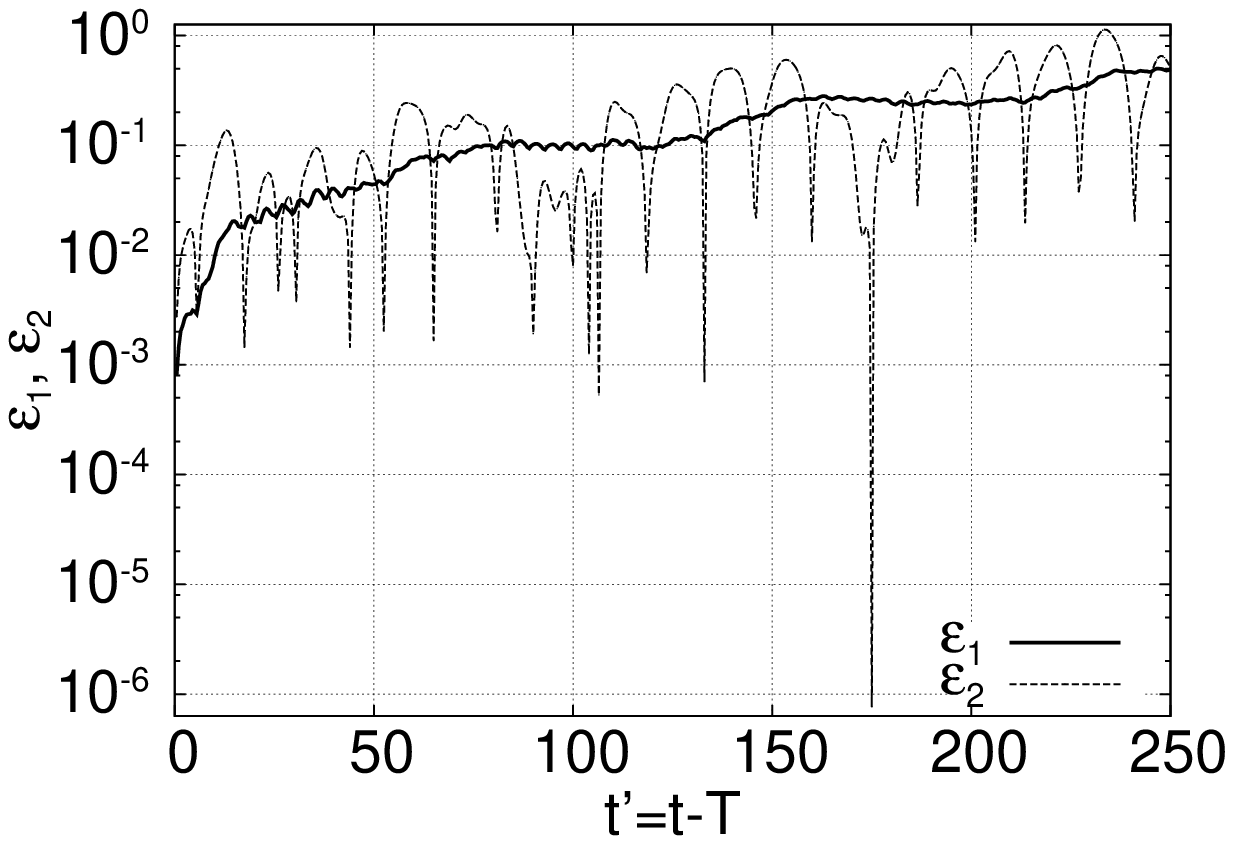}
\includegraphics[width=0.99\columnwidth,height=0.240\columnwidth]{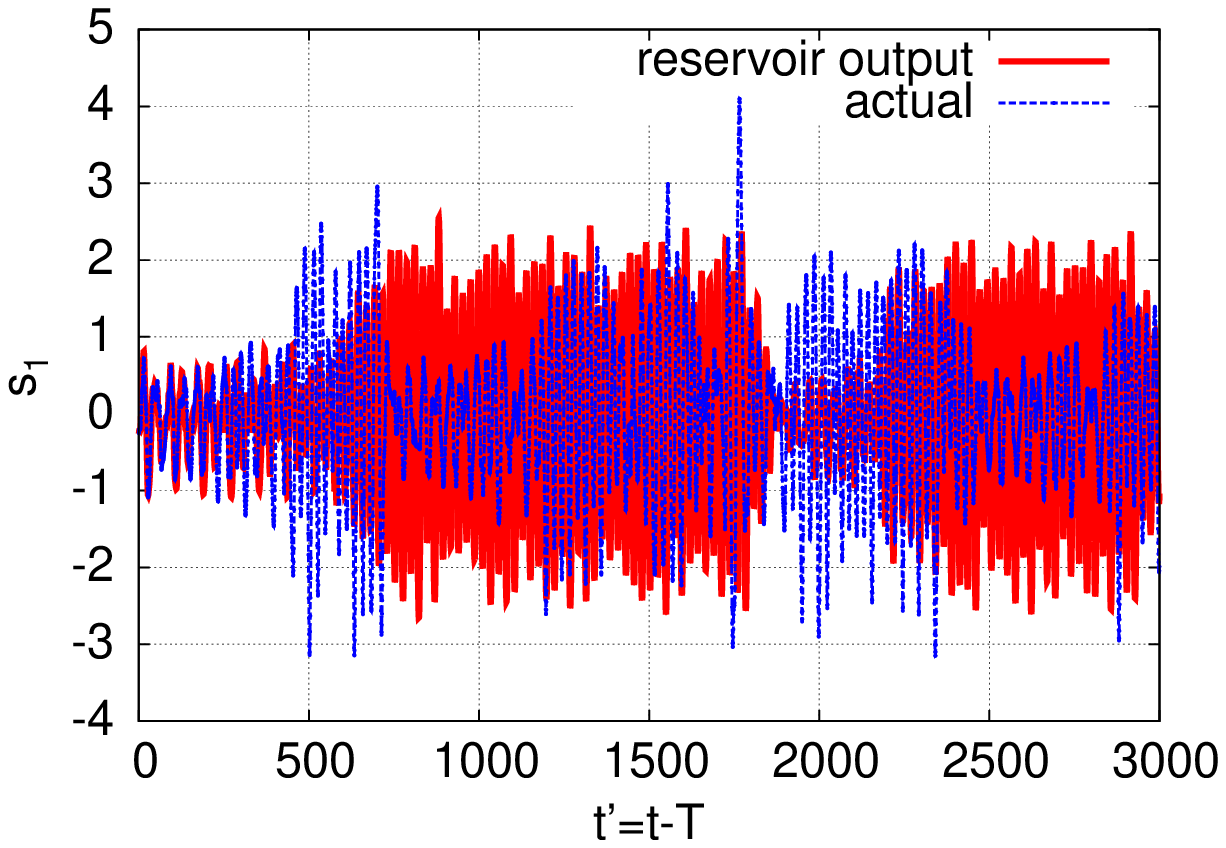}
\caption{ {\bf 
Inference of a time-series of the Reynolds number of a fluid flow.}
Time-series of $s_1=\tilde{R}_{\lambda}$ is inferred from the reservoir model 
in comparison with that of a reference data obtained by the direct numerical simulation of the Navier-Stokes equation 
(top left). 
The variable $t^\prime ~(= t-T>0)$ denotes the time after finishing the training phase at $t=T$. 
The inference errors $\varepsilon_1, \varepsilon_2$ defined by $\varepsilon_1(t)=|\mb{s}(t)-\hat{\mb{s}}(t)|$, and 
$\varepsilon_2(t)=|s_1(t)-\hat{s}_1(t)|=|\tilde{R}_{\lambda}(t)-\hat{\tilde{R}}_{\lambda}(t)|$ are shown to 
increase exponentially due to the chaotic property
(top right). 
In the bottom figure switching between laminar state with a small amplitude fluctuation and bursting state with a large amplitude fluctuation appear in an inferred time-series of $s_1=\tilde{R}_{\lambda}$,  
which are observed in the actual time-series. 
Remark that the model trajectory shows an intermittent behavior on a chaotic set, but after a long transient it will diverge eventually at 
$t^{\prime}\approx290000$. 
\allblack
}\label{fig:full-macro}
\end{figure}
\subsection{Evaluation of the model}
\begin{figure}[h]
\begin{center}
  \includegraphics[width=0.6\columnwidth,height=0.240\columnwidth]{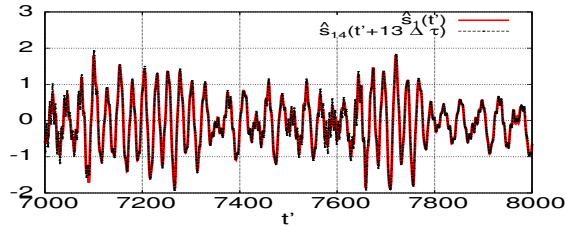}\\
  \caption{{\bf  Reproducing the delay property which is to be satisfied for the successfully inferred time-series $\hat{\bf s}$.}
  We observe that  for all values of $m=2,\cdots, 14$ and for most $t^{\prime}$, $\hat{s}_1(t^\prime)\approx\hat{s}_{m}(t^\prime+(m-1)\Delta\tau)$, 
  although the time-series of only $\hat{s}_1(t^\prime)$ and $\hat{s}_{14}(t^\prime+13\Delta\tau)$ $(7000\le t^{\prime}\le 8000)$ are shown. 
  The delay property is reproduced only from the training data. 
  }\label{fig:delay-time-series}
  \end{center}
\end{figure}

\begin{figure}[h]
\begin{center}
  \includegraphics[width=0.6\columnwidth,height=0.240\columnwidth]{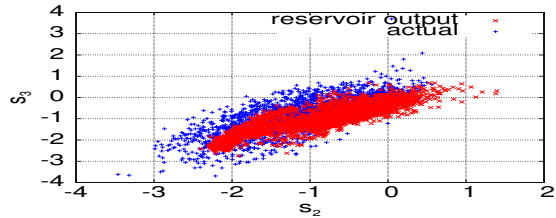}
  \caption{{\bf 
  Poincar\'e points on the plane $(s_2,s_3)$ along the trajectory $\Hat{\mb{s}}$ obtained from the reservoir model (red) and $\mb{s}$ from the Navier-Stokes equation (blue). }  
The time length of each trajectory is $90000$. The Poincar\'e section is defined by ${s}_{1}=0, ~d{s}_{1}/dt>0$.
Two sections are similar to each other, although a trajectory generated from the reservoir model does not cover some region of bursting states.
The figure suggests that each of the chaotic set is hyper-chaotic, that is the dimension of the unstable manifold is two or higher.}\label{poincare}
  \end{center}
\end{figure}

\begin{figure}[h]
\begin{center}
  \includegraphics[width=0.65\columnwidth,height=0.240\columnwidth]{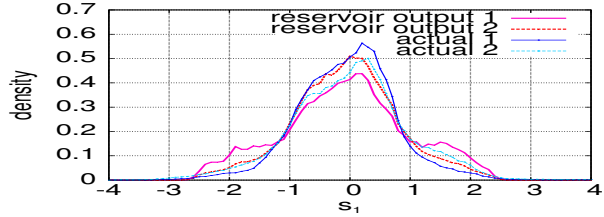}\\
  \caption{{\bf Density distributions generated from trajectories for a variable $s_1$ obtained from the constructed reservoir model (reservoir output) and from the direct numerical simulation of the Navier-Stokes equation (actual).}
  Each trajectory with a time-length 50000 has a different initial condition. The distributions are similar to each other in the sense that the peak is taken at $s_1\approx0.2$, and the distribution has relatively long tails. 
 }\label{distribution}
  \end{center}
\end{figure}

\begin{figure}[h]
\begin{center}
\includegraphics[width=0.99\columnwidth,height=0.240\columnwidth]{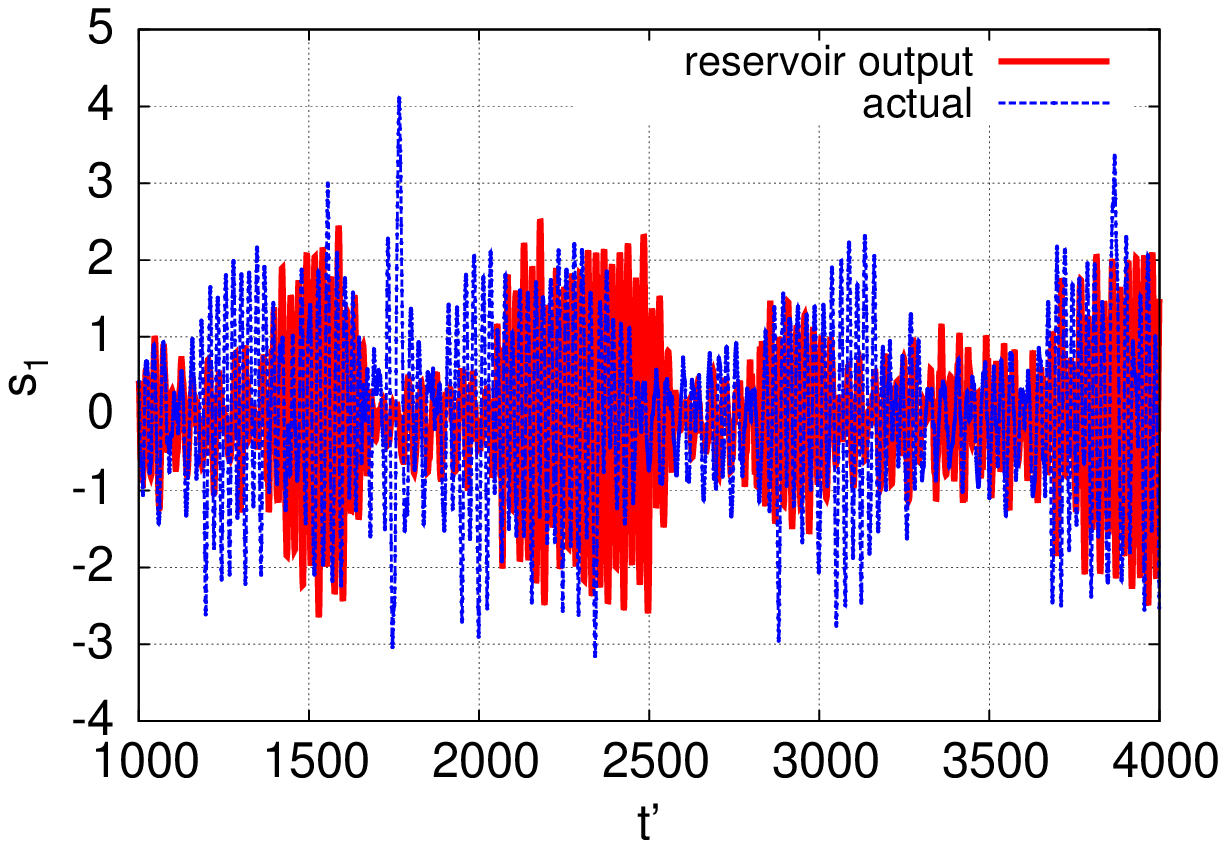}
  \includegraphics[width=0.6\columnwidth,height=0.240\columnwidth]{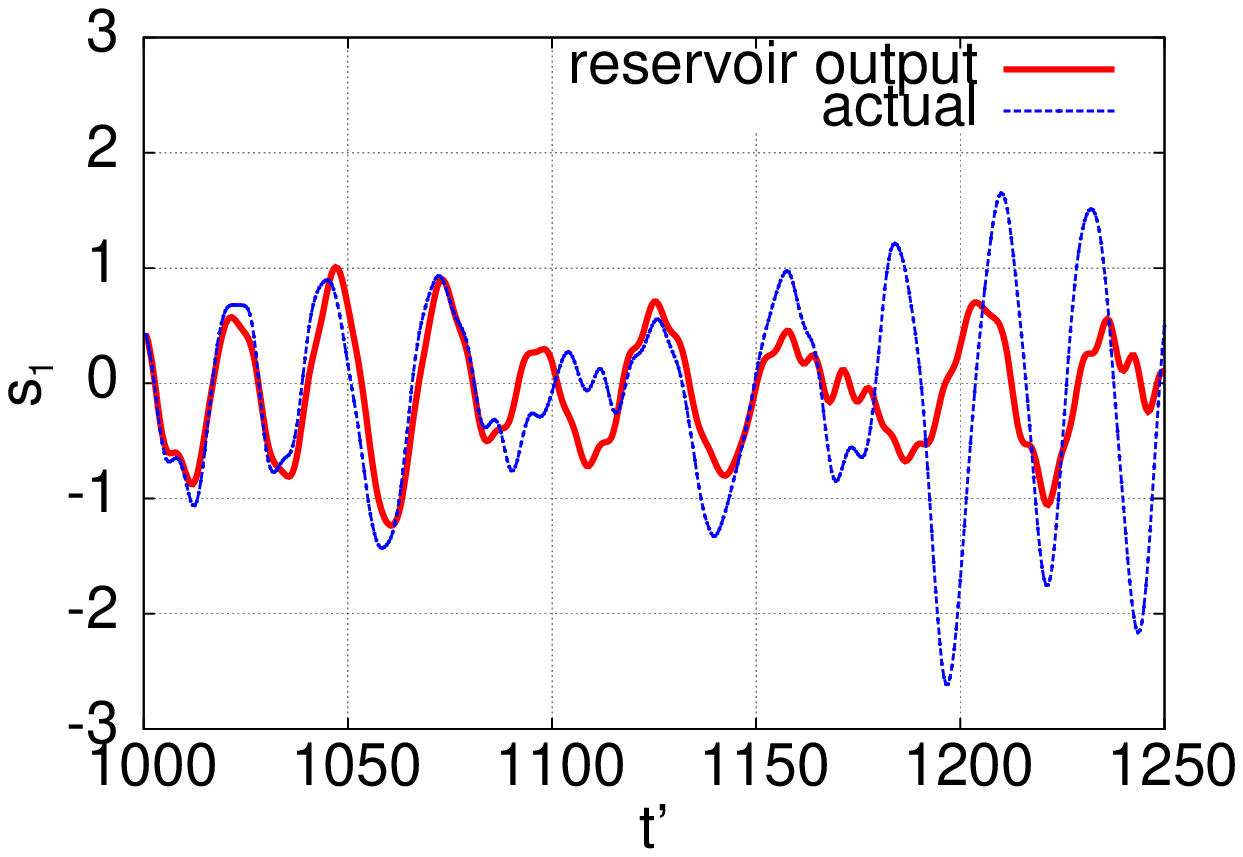}
  \caption{{\bf  
 Inference of a time-series of the Reynolds number for $t^{\prime}>T_{\text{out}}$ ($T_{\text{out}}=1000$) using the reservoir model constructed by using the training data for $t^\prime\le 0$ (see Fig.~\ref{fig:full-macro}). }
 We use the same $\mb{W}_\text{in},\mb{A},\mb{W}^*_\text{out}$ and $\mb{c}^*$ as those used for the model inferring the trajectory in Fig.~\ref{fig:full-macro}. 
 But we use  the time-series $s_1(t^\prime)$ for $T_{\text{out}}- T_1<t^\prime<T_{\text{out}}$ as an initial condition, where 
 $T_1$ is the transient time for the reservoir state vector $\mb{r}(t)$ to be converged. 
 \allblack 
In the top panel, switching between laminar and  bursting states is observed in the inferred trajectory. 
 The bottom panel is the enlargement of the top panel, and shows that the model has a predictability for $1000<t^\prime<1080$. 
 This means that the reservoir model constructed using the training data at a certain time-interval can become the model for another time-interval. 
}\label{fig:diff-inference}
  \end{center}
\end{figure}

\begin{figure}[h]
\begin{center}
  \includegraphics[width=0.327\columnwidth, height=0.2450\columnwidth]{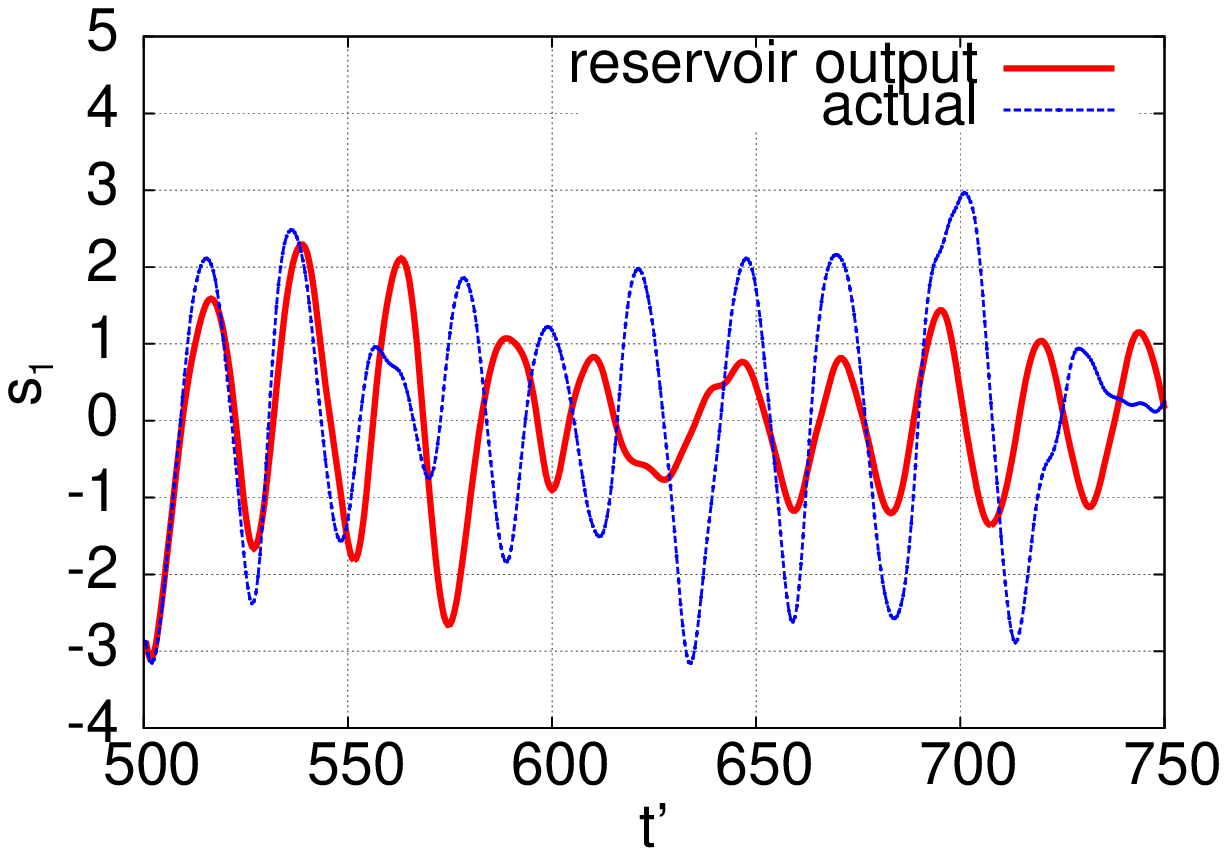}
  \includegraphics[width=0.327\columnwidth, height=0.2450\columnwidth]{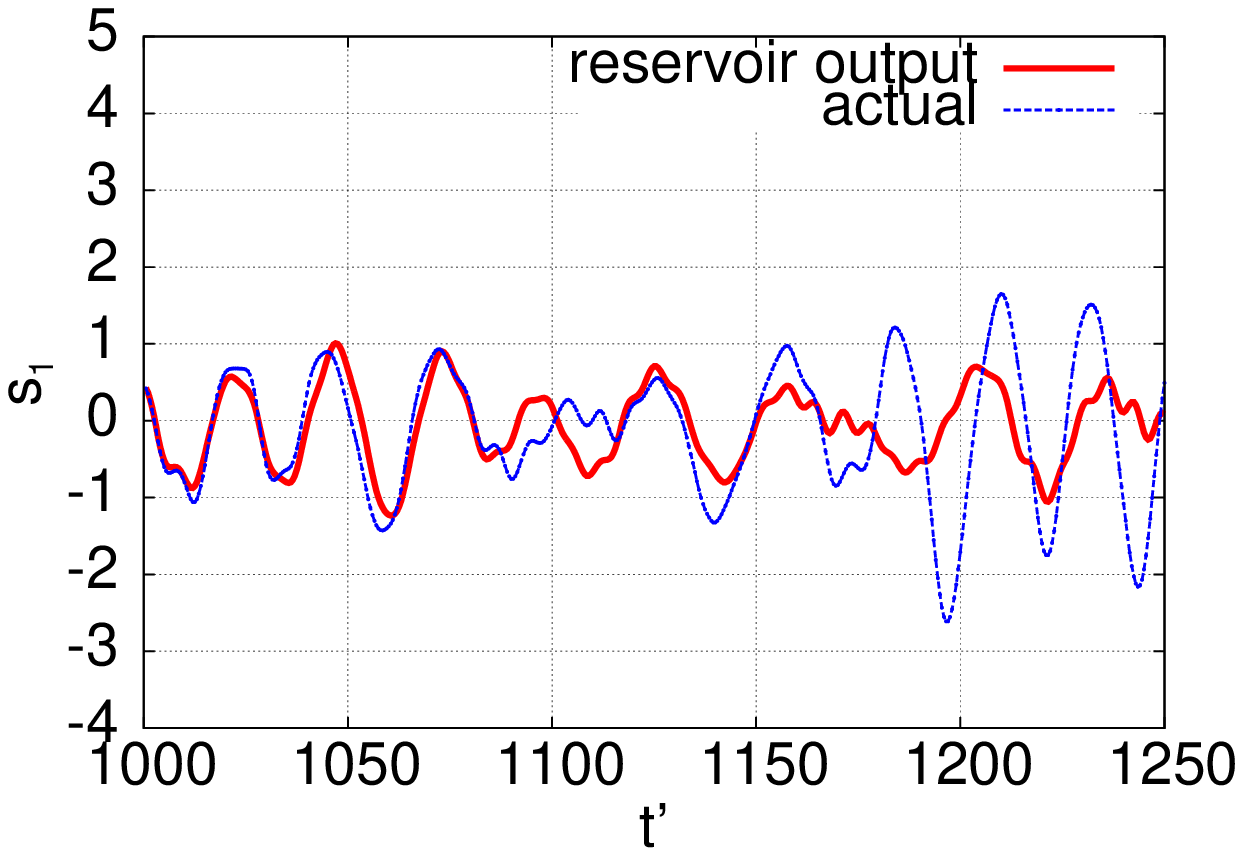}
  \includegraphics[width=0.327\columnwidth, height=0.2450\columnwidth]{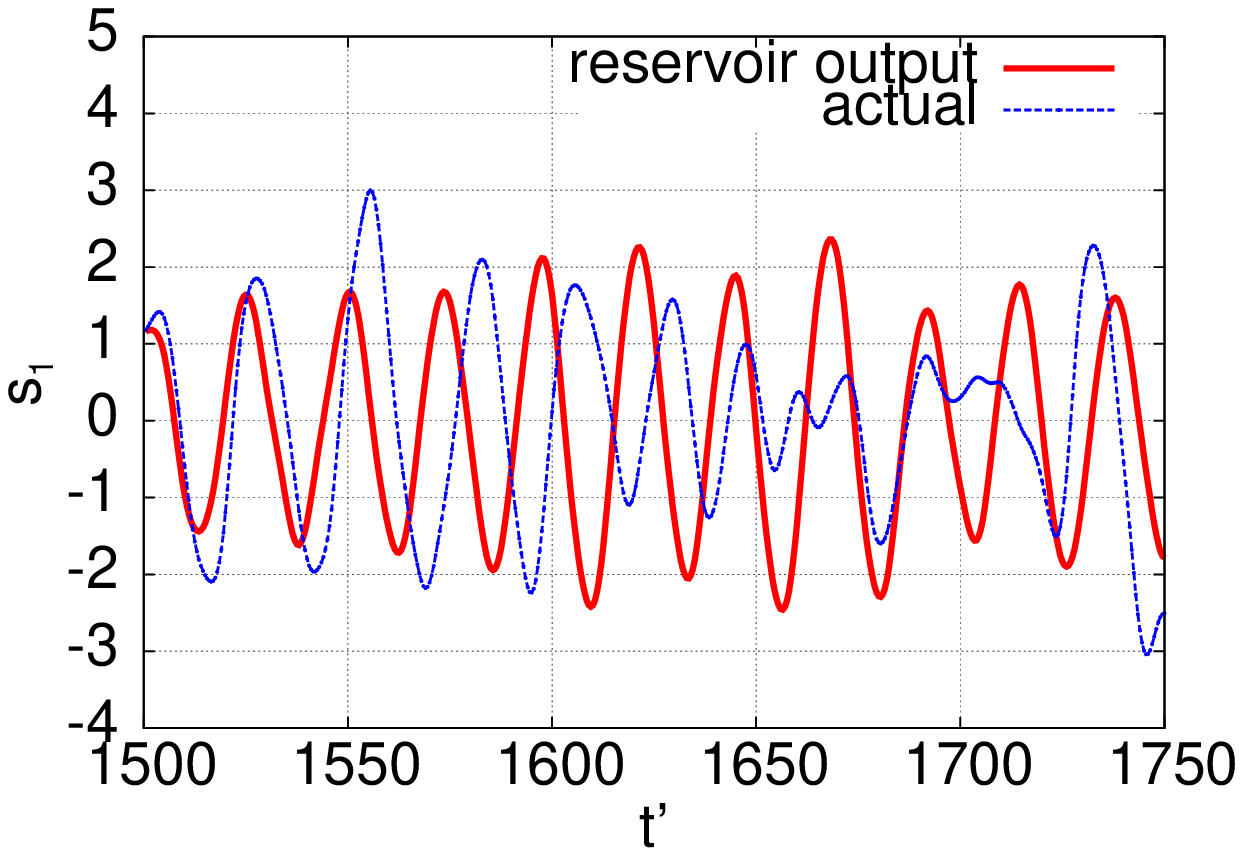}
  \includegraphics[width=0.327\columnwidth, height=0.2450\columnwidth]{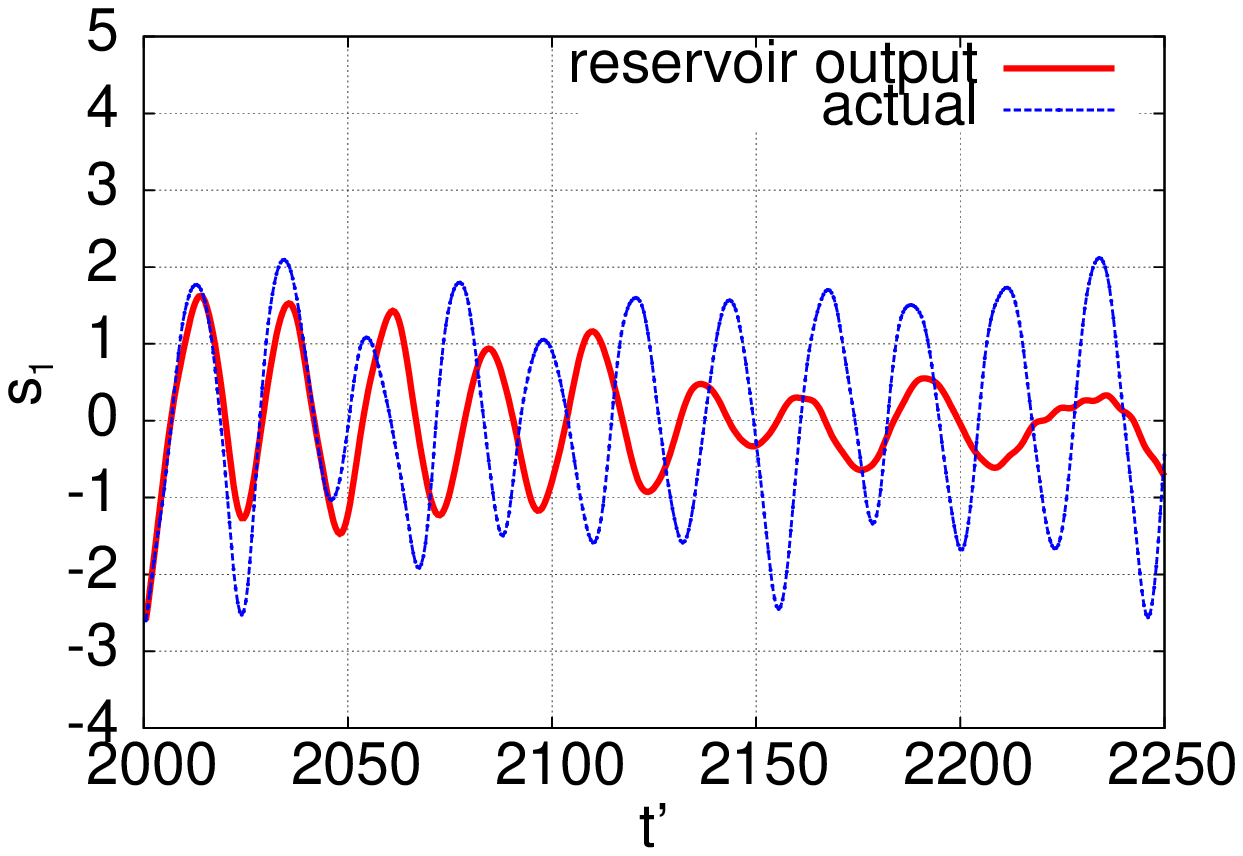}
  \includegraphics[width=0.327\columnwidth, height=0.2450\columnwidth]{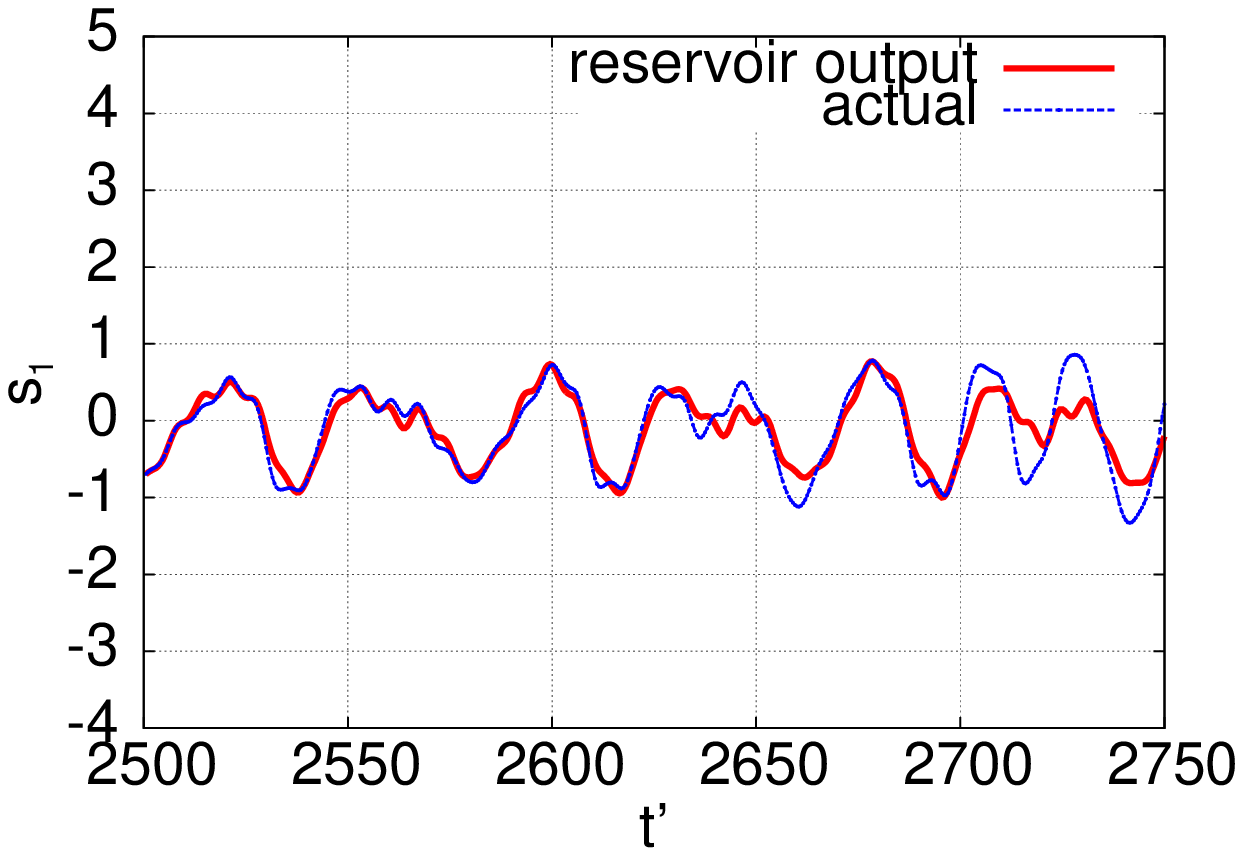}
  \includegraphics[width=0.327\columnwidth, height=0.2450\columnwidth]{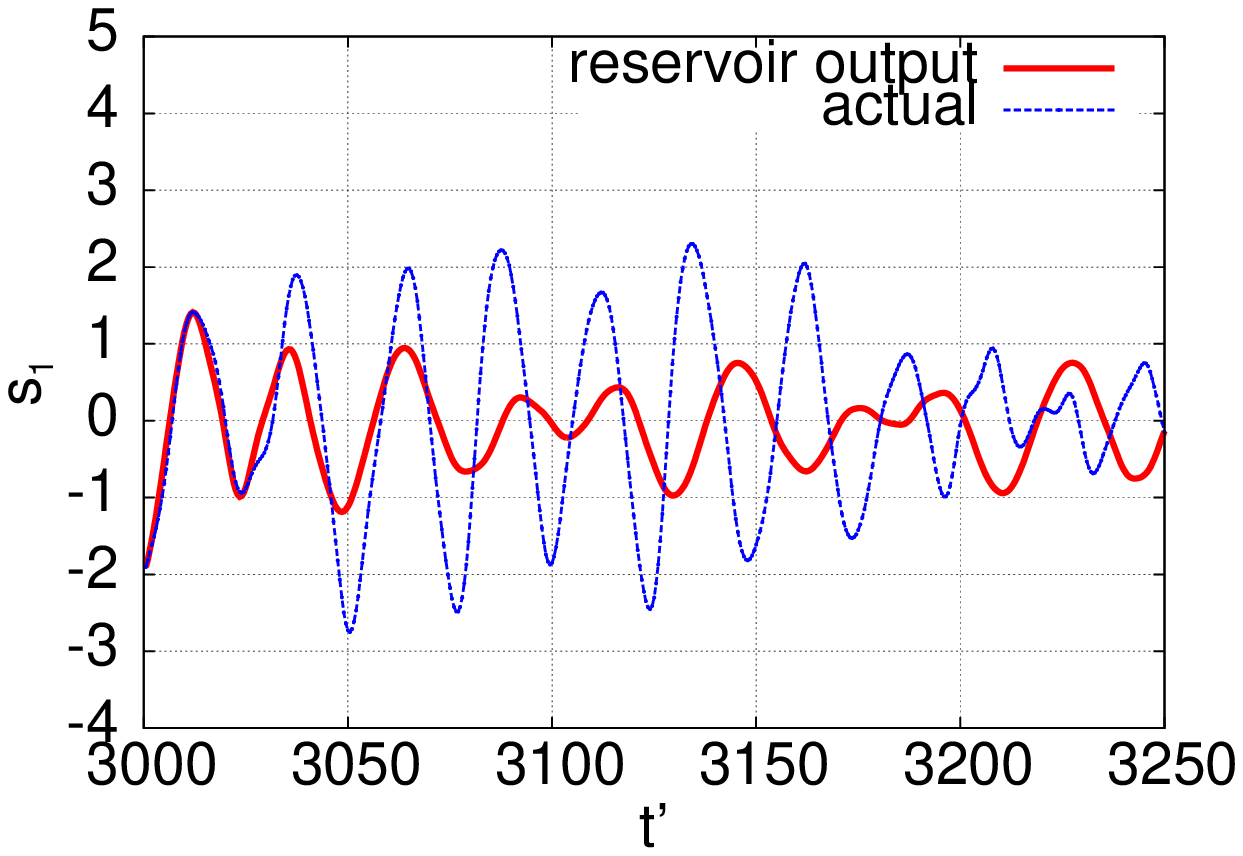}
  \includegraphics[width=0.327\columnwidth, height=0.2450\columnwidth]{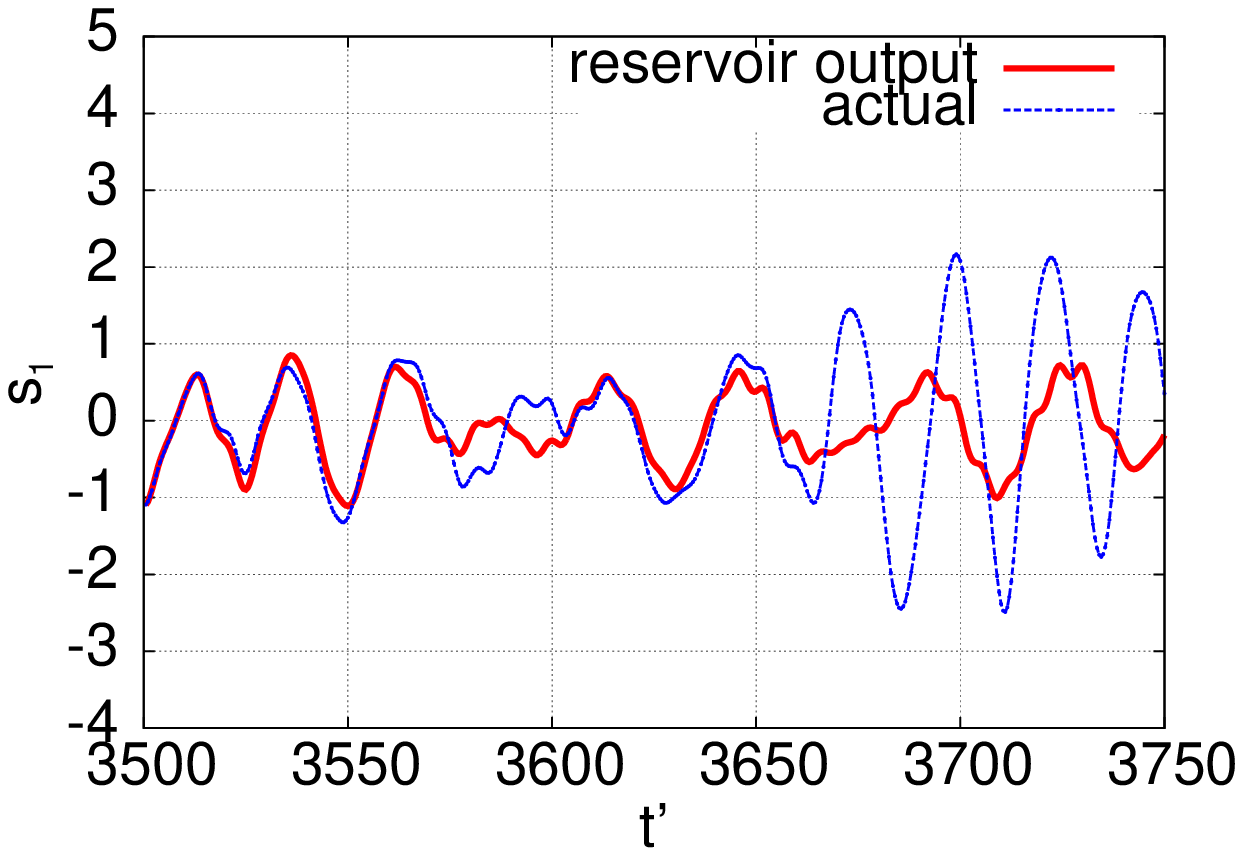}
  \includegraphics[width=0.327\columnwidth, height=0.2450\columnwidth]{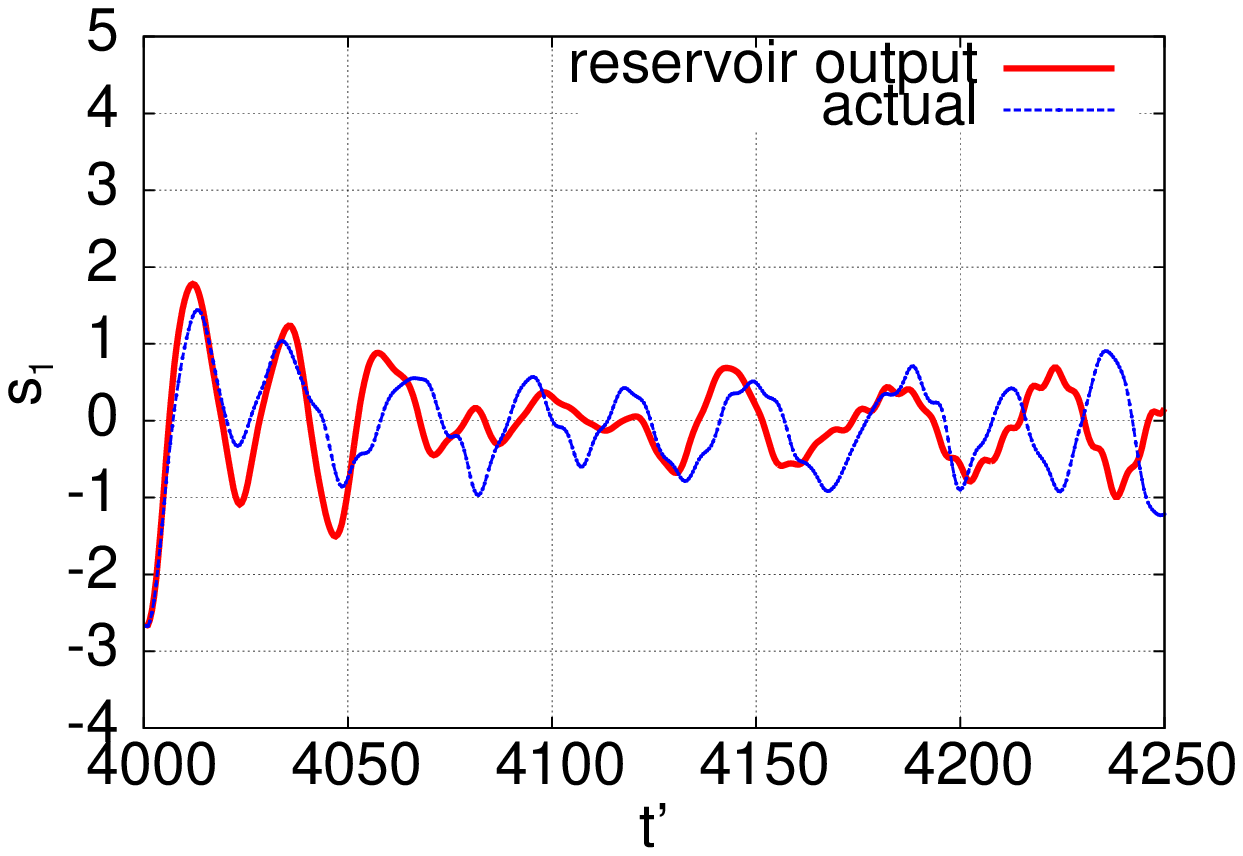}
  \includegraphics[width=0.327\columnwidth, height=0.2450\columnwidth]{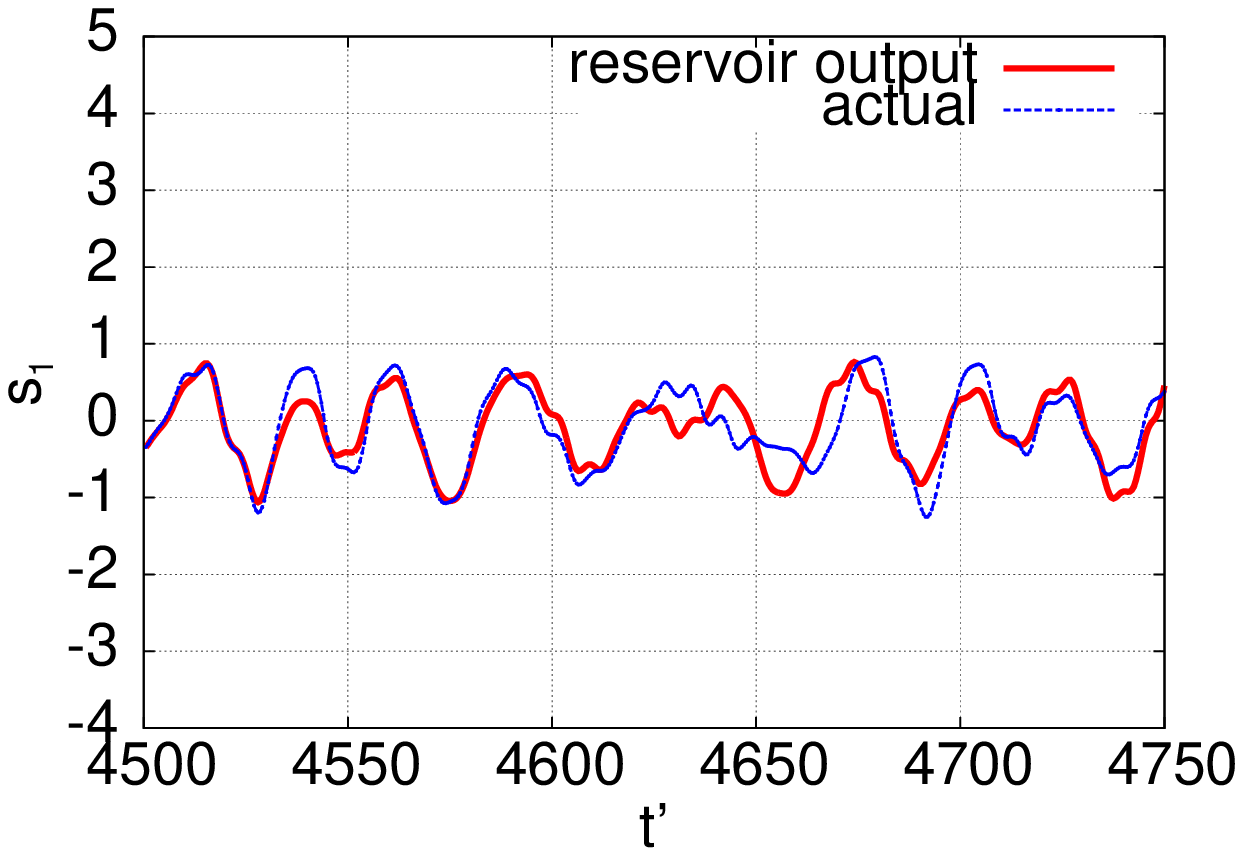}
  \includegraphics[width=0.327\columnwidth, height=0.2450\columnwidth]{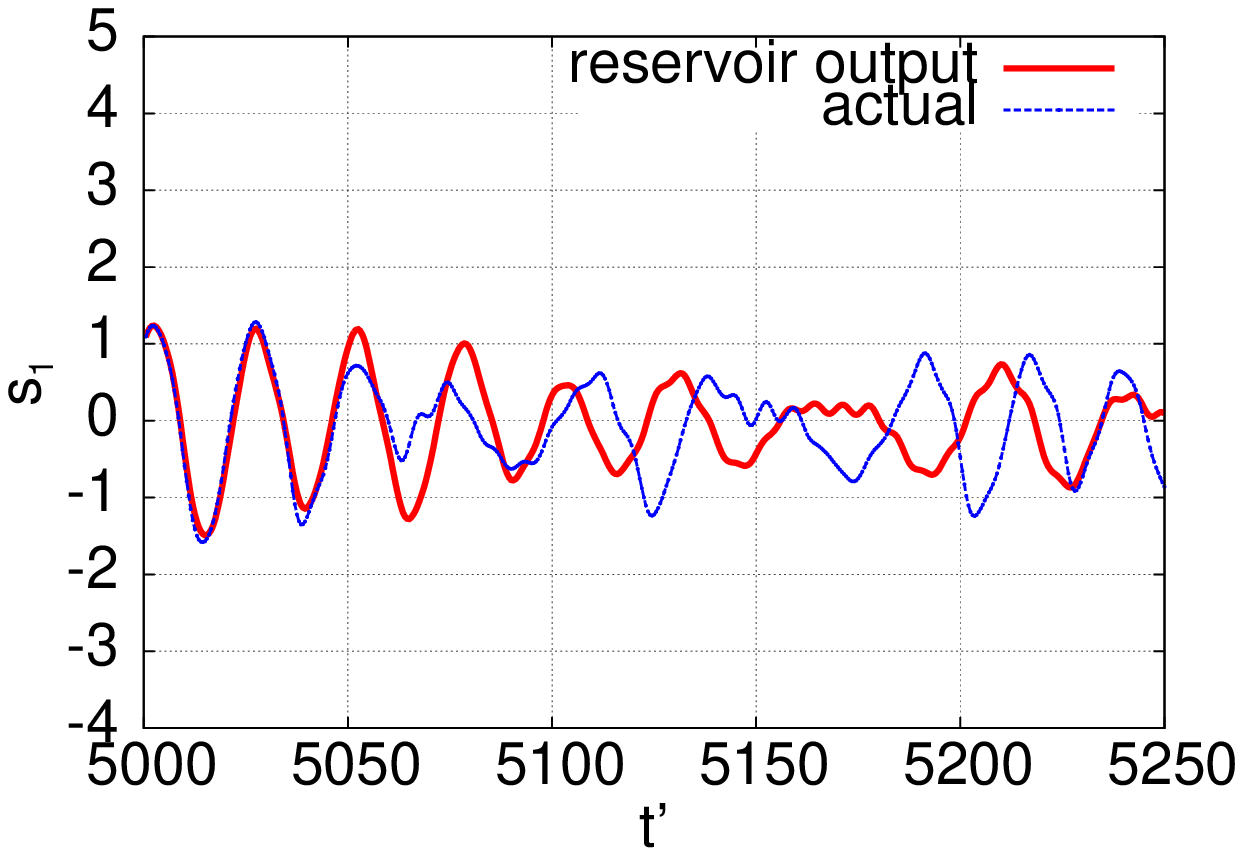}
  \includegraphics[width=0.327\columnwidth, height=0.2450\columnwidth]{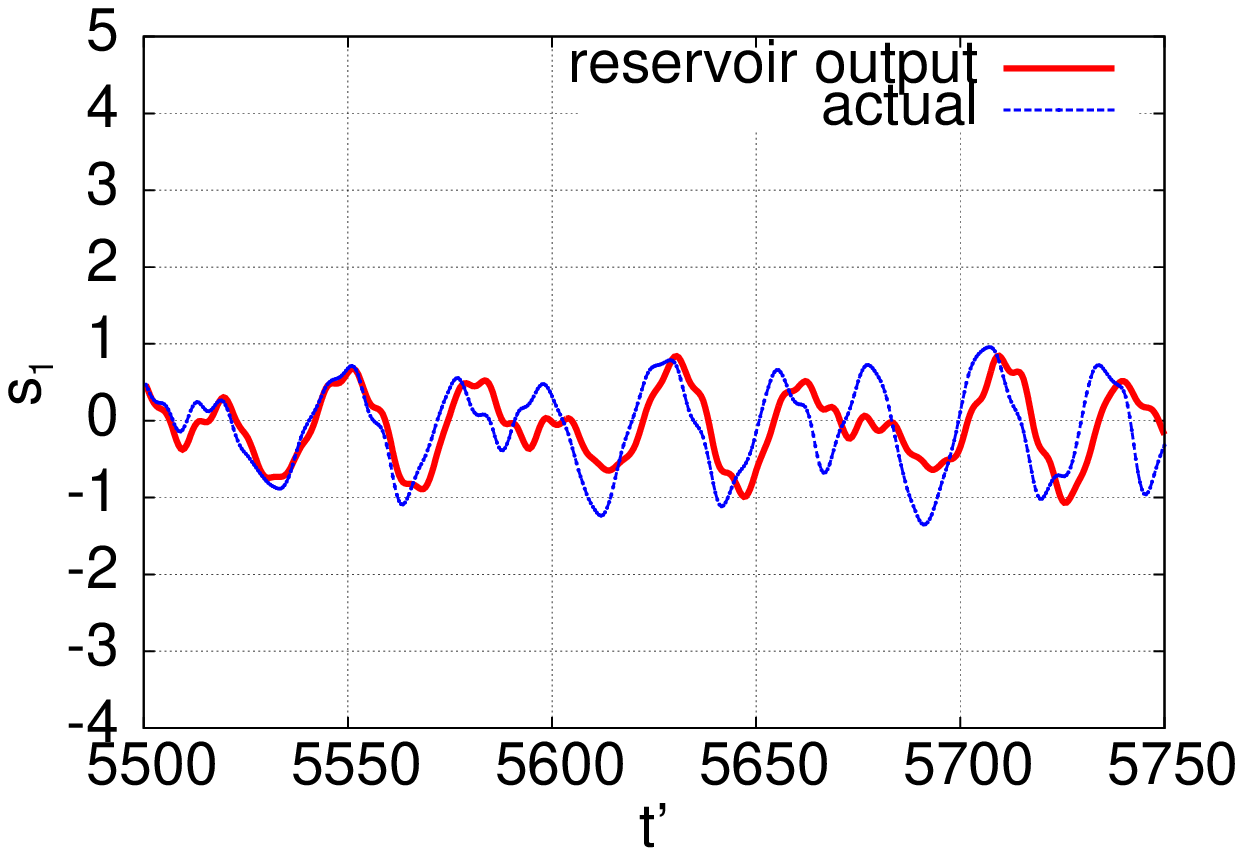}
  \includegraphics[width=0.327\columnwidth, height=0.2450\columnwidth]{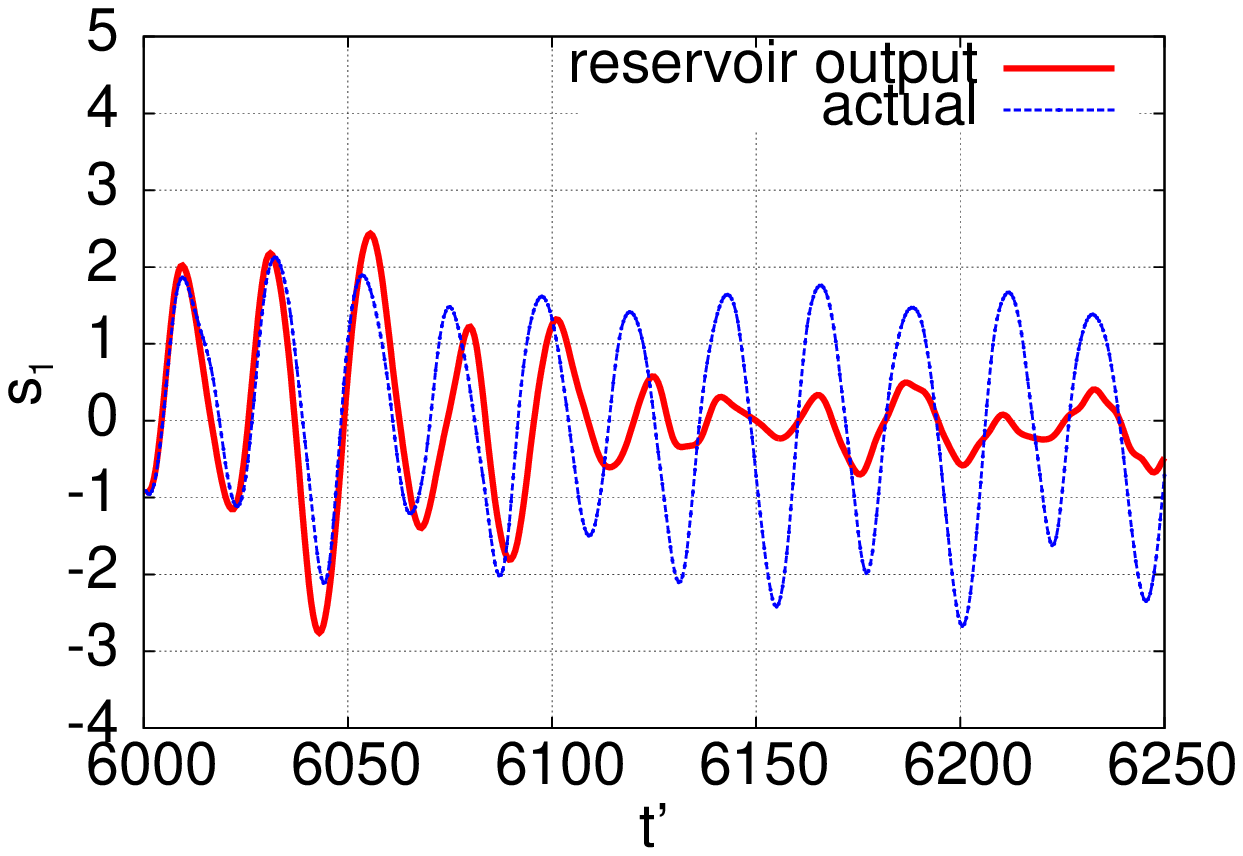}
  \caption{{\bf  Inference of time-series of the Reynolds number in many time-intervals $T_{\text{out}}<t^{\prime}<T_{\text{out}}+250$ ($T_{\text{out}}=500, 1000, \cdots, 6000$) using the same reservoir model constructed by using the training data for $t^\prime\le 0$ (see Fig.~\ref{fig:full-macro} and \ref{fig:diff-inference}.)}
As in Fig.~\ref{fig:diff-inference},  we only change the initial condition for each case, while the model is fixed after 
the appropriate choice of $\mb{W}_\text{in},\mb{A},\mb{W}^*_\text{out}$ and $\mb{c}^*$ is determined 
by using the training data for $t^{\prime}<0$. 
Each panel shows that the time-series inference is successful by using  the same model, although the 
time-interval for the successful inference is limited to a relatively short time especially for the bursting phase with large fluctuations 
maybe due to the high instability.
These panels  suggest that the model can describe the dynamics of the Reynolds number. 
}\label{fig:mult-diff-inference}
  \end{center}
\end{figure}

We evaluate the constructed reservoir model for the Reynolds number from several points of view 
by comparing its property with that of the actual data obtained from the direct numerical simulation of Navier-Stokes equation. \\
{\bf Time-series.}
We confirm that an inference of a time-series of the Reynolds number $s_1=\tilde{R}_{\lambda}$ is successful for some time after finishing the training phase. 
The time-series of the inferred variable $\hat{s}_1=\hat{\tilde{R}}_{\lambda}(t)$ ($t>T$) is shown with the actual data 
$\hat{s}_1={\tilde{R}}_{\lambda}(t)$ obtained from the direct numerical simulation of Navier-Stokes equation in the top left panel of Fig. \ref{fig:full-macro}. 
The failure in the long-term time-series inference is inevitable just due to the sensitive dependence on the initial condition of a chaotic property of the fluid flow. 
The two types of errors between the inferred value and the actual one are shown in the top right panel of Fig. \ref{fig:full-macro}. 
Moreover, the long-time behavior of $\hat{\tilde{R}}_{\lambda}(t)$ is shown in the bottom panel, 
which has qualitatively similar intermittent behaviors to the actual one,  intermittent switching between the state of low amplitude fluctuations (laminar state) and the state of high amplitude fluctuations (burst state).
\\
{\bf Delay Property.}
\allblack
As we employ the delay-coordinate vector for input and output variables of the reservoir computing (eqs.~\eqref{eq:delay},\eqref{eq:delay2}), 
the relation $s_1(t) = s_{m}(t+(m-1)\Delta \tau)$ holds for any $m ~(m=2,\cdots,M)$ during the training phase.
The corresponding relation should also be satisfied in the inference phase.
We show the time-series $\hat{s}_1(t)$ and $\hat{s}_{14}(t+13\Delta \tau)$ in Fig. \ref{fig:delay-time-series},
which satisfies the relation $\hat{s}_1(t) \approx \hat{s}_{14}(t+13\Delta \tau)$. 
We can confirm that for almost all $t$ the relation $\hat{s}_1(t)\approx \hat{s}_{m}(t+(m-1)\Delta \tau)$ is satisfied for any $m$. 
The results imply that our reservoir computing successfully learns the delay property only through training such data.
\\
{\bf Poincar\'e plane.}
We investigate the chaotic set computed from a model trajectory to see whether the inferred chaotic set mimics the actual one.
For its purpose we describe the Poincar\'e plane 
in comparison with that computed from a trajectory of the direct numerical simulation of the Navier-Stokes equation with the same time length in the Fig. ~\ref{poincare}. 
Although the sections are similar to each other, but they are not very close to each other. This may be because the length of the 
intermittent trajectory is not enough to cover various regions especially in the bursting state.
\\
{\bf Distribution.}
Density distributions computed from two inferred trajectories $\{ \hat{s}_1(t) \}$ and those from two actual trajectories 
$\{ {s}_1(t) \}$ are shown in Fig. ~\ref{distribution}. 
We can observe that the distributions computed from trajectories of time lengths 5000 are fluctuating, but the inferred distributions
 seem to have similar properties to the actual distributions. 
Relatively large fluctuations in distributions for 
 $|\hat{s}_1(t) |>1$ should be due to the intermittency.
%
%
%
%
%
\\
{\bf The reservoir model can be used to infer time-series of another time-interval.} 
We obtained a model just by training the data and it enables us to infer short-time behavior, the shape of an attractor and the density distribution. 
Here we confirm that the model constructed using a certain training data has the ability to infer a short-time behavior of the Reynolds number for the totally different time-interval. 
In Fig.~\ref{fig:diff-inference}, the inferred time-series is shown in comparison with the actual one.
For this inference we use the same reservoir model as is used in Fig.~\ref{fig:full-macro}, and only change the initial condition.
This figure supports the accuracy of the constructed reservoir model. 
%
%
In Fig. \ref{fig:mult-diff-inference}, by using the same model the inference of time-series of the Reynolds number
 in many different time intervals are shown. For each time-interval, we confirm that the short time inference is successful. 
This implies that the obtained model can describe the dynamics of the Reynolds number. 
Remark that the top middle panel in Fig. \ref{fig:mult-diff-inference} corresponds to Fig. \ref{fig:diff-inference}.
\allblack
\clearpage
%

\section{\bf Choice of delay-coordinate.}\label{delay}
\begin{figure}[b]
\begin{center}
  	\includegraphics[width=0.48\columnwidth,height=0.240\columnwidth]{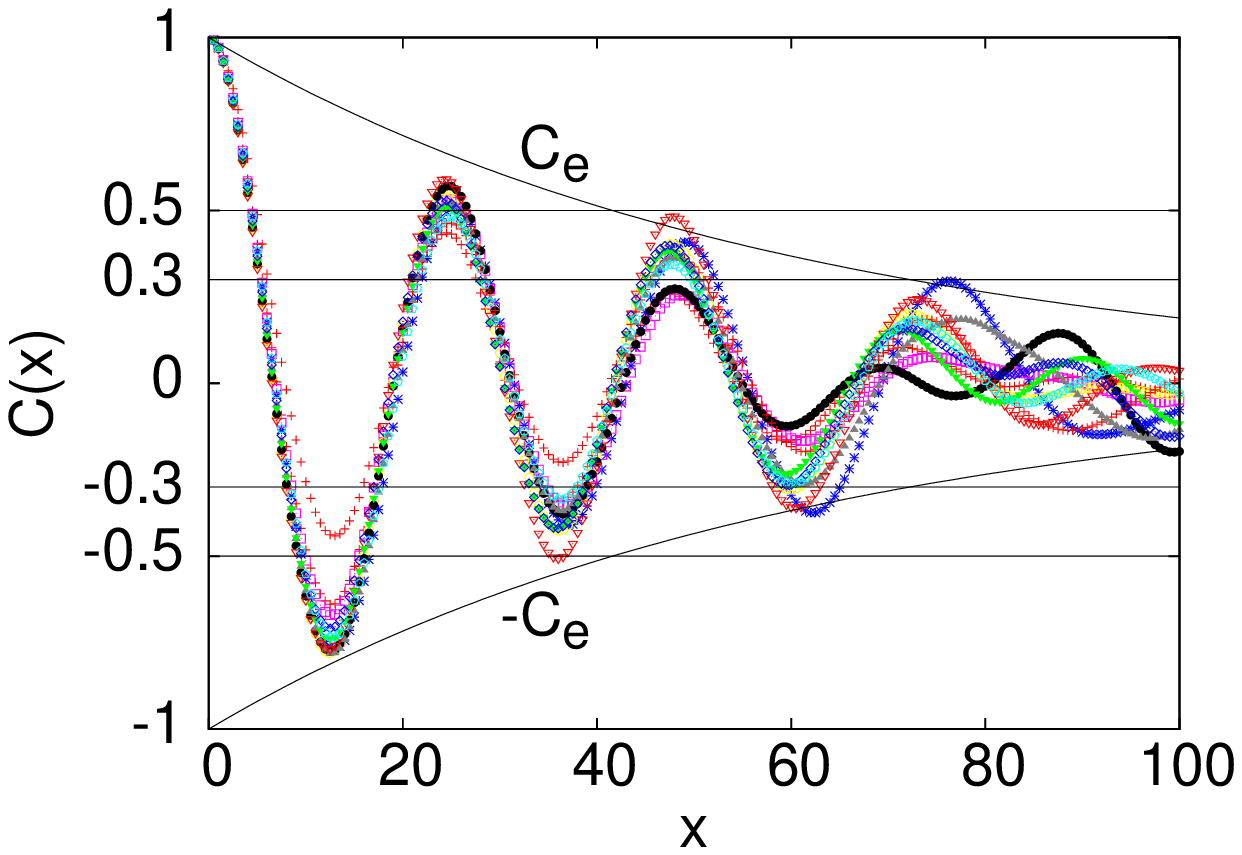} 
  	\includegraphics[width=0.48\columnwidth,height=0.240\columnwidth]{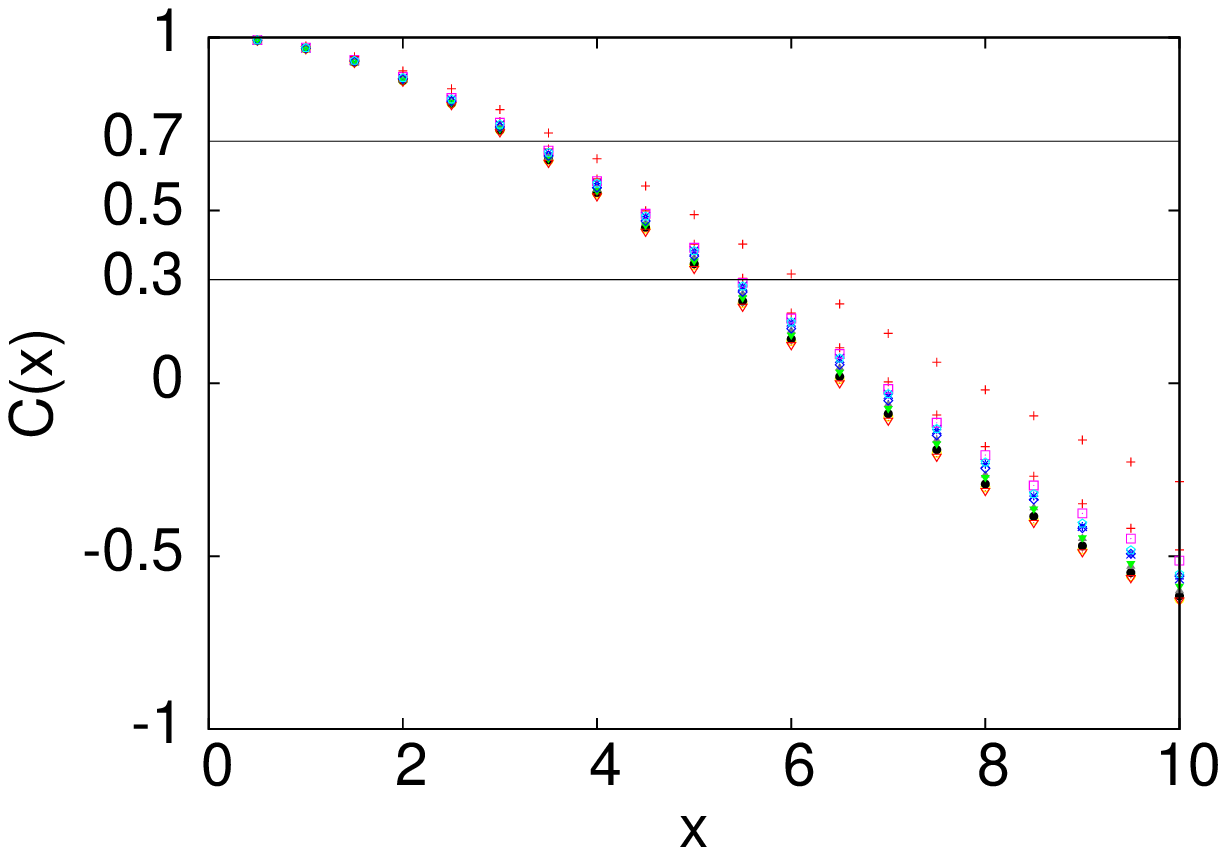}
\vspace{5mm}
  \caption{{\bf 
  Auto-correlation function $C(x)$ for a trajectory $\{R_{\lambda}(t)\}$
  with respect to the value of time-delay $x$ (left), and its enlarged figure (right).}
  Auto-correlation function $C(x)$ is shown together with 
  the straight lines $\pm0.3,\pm0.5$ (left panel), and $0.3,0.7$ (right panel).
  Each of the different colors represents $C(x)$ computed from a trajectory from a different initial condition with time-lengths 5000. 
  The difference is mainly due to the intermittent property of the dynamics.
  In the left panel the envelope $C_e(x)(=\exp(-x/60))$ is shown to go below 0.5 when $x \approx 40$,  and also go below 0.3 when $x \approx 75$.
 From the right panel $C(x)$ is shown to go below $0.7$ at the first time, when $x \approx  3.0$, 
and go below $0.3$ at the first time, when $x \approx 5.0$.
}
  \label{correlation}
  \end{center}
\end{figure}

\allblack
We use an $M$-dimensional delay-coordinate vector with a delay-time $\Delta \tau$ (eqs.~\eqref{eq:delay},\eqref{eq:delay2}) as input and output variables $\mb{u}$ and $\mb{s}$ in eq.~\eqref{eq:input}.
In this section we investigate the appropriate choice of time-delay $\Delta \tau$ and the dimension $M$.\\
{\bf Time-correlation.}
The auto-correlation function $C({x})$ along a trajectory $\{R_{\lambda}(t))\}$ with respect to the time-difference $x$ is computed by 
\begin{align}
C({x}) = \dfrac{\dfrac{1}{J}\displaystyle\sum_{j=0}^{J-1}(R_{\lambda}(t_0+j\Delta t^*)-\bar{R}_{\lambda})(R_{\lambda}(t_0+j\Delta t^*+x)-\bar{R}_{\lambda})}
	{
	\sqrt{\dfrac{1}{J}\displaystyle\sum_{j=0}^{J-1}(R_{\lambda}(t_0+j\Delta t^*)-\bar{R}_{\lambda})^2}
	\sqrt{\dfrac{1}{J}\displaystyle\sum_{j=0}^{J-1}(R_{\lambda}(t_0+j\Delta t^* +x)-\bar{R}_{\lambda})^2}
	},
\end{align}
where 
$\bar{R}_{\lambda}$ is the time average of $R_{\lambda}(t)$, $\Delta t^*$ is the time step of the discrete trajectory, and 
$t_0$ is an initial  time of a trajectory. 
In Fig.~\ref{correlation}, we show the auto-correlation function $C(x)$ for a trajectory $\{R_{\lambda}(t)\}$ 
with respect to the delay-time $x$.
It is observed from Fig.~\ref{correlation}~(right) that as $x$ increases from 0, 
$C(x)$ goes below 0.7 and 0.3 when $x\approx 3.0$ and $5.0$, respectively. 
\\
\indent The observation suggests that the value of the delay-time $\Delta \tau$ is to be chosen around $3.0$-$5.0$. 
If $\Delta \tau< 3.0$, the consecutive two components of a delay-coordinate vector in \eqref{eq:delay}, 
$R_{\lambda}(t)$ and 
$R_{\lambda}(t-\Delta \tau)$  behave too similarly, and if $\Delta \tau> 5.0$, the consecutive two components behave too differently, 
and some dynamics to be captured may be missing. 
\begin{table}[h]
\normalsize
(a)~$(e_{60},e_{90})=(0.14,0.30)$
		\begin{tabular}{|l|r|r|r|r|r|r|r|r|r|r|r|r|r|r|} 
			\hline  	
           		 $\Delta \tau$ $\backslash$ $M$ &10&11&12&13&14&15&16&17&18&19&20 \\ \hline
  			3.0&0&0&0&0&0&1&19&24&\blue{\bf43}&37&27\\ \hline
  			3.5&0&0& 0&11&20&28&\blue{\bf57}&48&21&11&7\\ \hline
  			4.0&0&3&18&43&\underline{\red{\bf107}}&59&21&14&2&4&5 \\ \hline
  			4.5&3&14&43&\blue{\bf54}&21&15&8&1&1&1&0 \\ \hline
  			5.0&10&24&\blue{\bf26}&19&9&1&1&1&0&0&0\\ \hline
 \end{tabular}
(b)~$(e_{60},e_{90})=(0.13,0.17)$
			\begin{tabular}{|l|r|r|r|r|r|r|r|r|r|r|r|r|r|} 
			\hline  	
           		$\Delta \tau$ $\backslash$ $M$ &10&11&12&13&14&15&16&17&18&19&20 \\ \hline
  			3.0&0&0&0&0&0&0&3&6&\blue{\bf10}&8&4\\ \hline
  			3.5&0&0& 0&2&3&5&\blue{\bf6}&4&1&3&1\\ \hline
  			4.0&0&0&2&8&\underline{\red{\bf14}}&10&1&4&1&0&1 \\ \hline
  			4.5&1&1&8&\underline{\red{\bf14}}&1&0&1&0&0&0&0 \\ \hline
  			5.0&2&4&\blue{\bf6}&\blue{\bf6}&3&0&1&0&0&0&0\\ \hline
 \end{tabular}
 		\caption{{\bf The number of successful trials for each choice of the delay-time $\Delta \tau$ and the dimension $M$ of the delay-coordinate.} 
 		The matrices $\mb{A}$ and $\mb{W}_{\text{in}}$ are chosen randomly, and the number of successful cases are counted. 
See Table.~\ref{tab:parameter} for the parameter values.
 		We say the inference is successful, if the three conditions (i)(ii)(iii) in ~\eqref{eq:defErrer} hold,
where the criteria $(e_{60},e_{90})$ are set as (a)$(0.14,0.30)$ and (b)$(0.13,0.17)$. 		
For each set of values $(\Delta \tau,M)$ we tried 8160 cases of $\mb{A}$ and $\mb{W}_\text{in}$.
For each value of $\Delta\tau$, the best choice of $M$ is identified by the bold number(s) (blue), and 
the best among each criterion is identified by the underlined bold number(s) (red).
 		}
 		\label{tab:parameter-delay}
\end{table}


{\bf Delay-time and dimensions.}
Based on the above implication about the auto-correlation function in Fig.~\ref{correlation}, 
 we investigate the effective delay-time $\Delta \tau$ and dimensions $M$. 
We focus on the delay-time $\Delta \tau \approx 3.0$-$5.0$ in Table \ref{tab:parameter-delay}. 
We infer time-series of the Reynolds number $R_{\lambda}(t)$ (actually its normalized value $\tilde{R}_{\lambda}(t)$) using the procedure in Sec.~\ref{reservoir} 
by employing the delay-coordinate in eq.~\eqref{delay}.
We tried 8160 cases for each set of parameters $(\Delta \tau,M)$, for which matrices $\mb{A}$ and  $\mb{W}_\text{in}$ are chosen randomly, and the number of successful cases are counted in TABLE \ref{tab:parameter-delay}. 
We say that the inference of $s_1(t^\prime)$ ($t^\prime = t-T>0$)  is successful if the conditions 
\begin{align}
\begin{cases}\label{eq:defErrer}
 \text{(i) the time average along } |\hat{s}_1(t^\prime)|<3 \text{ for } t^\prime \le 3000,\\
 \text{(ii) the error }\varepsilon_2(t^\prime)=|s_1(t^\prime)-\hat{s}_1(t^\prime)|=|\tilde{R}_{\lambda}(t^\prime)-\hat{\tilde{R}}_{\lambda}(t^\prime)|<e_{60} \text{ for all }t^\prime\le 60, \\
 \text{(iii) the error }\varepsilon_2(t^\prime)<e_{90} \text{ for all }t^\prime\le 90, 
\end{cases}
\end{align}
hold, where the criteria $(e_{60},e_{90})$ are set as (a)$(0.14,0.30)$ and (b)$(0.13,0.17)$. 
Remark that the condition (i) is given so as to get rid of a candidate which diverges within a short time, 
as $|s_1(t^\prime)|<3$ for almost all $t$ even in the bursting region. 
For each case we use the same training data as in Fig.~\ref{fig:full-macro}.

It is observed that the delay-time $\Delta \tau$ and the dimension $M$ of the delay-coordinate are chosen so that  $\Delta \tau\approx 4.0$-$4.5$, and $M\Delta \tau\approx 55$-$60$, 
which correspond to $C(\Delta \tau)\approx0.45$-$0.55$ and its envelope $C_e(M \Delta \tau) \approx0.35$-$0.40$, respectively (see the left panel of Fig. \ref{correlation}  for   the envelope $C_e$). 
For $\Delta\tau=4.0$ and $M=14, 15$ by computing 16 times more cases,  we confirmed that  the rate of successful trials does not change much.
In addition, even when we change the value of $N$ such as $1000$ or $3000$, 
we obtain almost the same results. 
\allblack
\clearpage

\section{Summary and discussion}\label{summary}
By training a time-series data of a macroscopic quantity the Reynolds number of a fluid flow, we construct a closed form system  describing its intermittent behavior between laminar and bursting states.
For the model construction, we do not use the knowledge of a physical process. 
We evaluate the obtained model in many ways. In particular, the model is confirmed to have a time-series predictability in many time intervals.\\
\indent In order to construct a model from a scalar time-series data, we introduce a time-delay coordinate. 
From our investigations, the time-delay should be chosen to be the lowest value $\Delta\tau (>0)$
	so that the auto-correlation function $C$ is $0.45<C(\Delta\tau)<0.55$ at the first time, 
	and that the dimension $M$ of the delay-coordinate  should be chosen so that the envelope $C_e$ of the auto-correlation function 
	$C$ is $0.35<C_e(M\Delta\tau)<0.40$.\\  
\indent It should be remarked that the obtained reservoir model has a chaotic set on which a trajectory approximates the actual one, 
but the set is not an attractor. 
This may be due to the lack of a training data, especially in the bursting state. The clarification is remained as a future study.
\allblack

\allblack

\section*{Acknowledgements}
KN was supported by the Leading Graduate Course for Frontiers of Mathematical Sciences and Physics (FMSP) at the University of Tokyo.
YS was supported by the JSPS KAKENHI Grant No.17K05360 and JST PRESTO JPMJPR16E5.
Part of the computation was supported by the Collaborative Research Program for Young $\cdot$ Women Scientists of ACCMS and IIMC, Kyoto University.


\bigskip
\bigskip
\bibliographystyle{apalike}
\bibliography{reservoir_bibliography,qr_bibliography}


\end{document}